\documentstyle[prl,aps,multicol,epsf]{revtex}
\begin{document}
\draft
%--------------------------------------------------------------------------
% Definition of title page:
\title{
        Euclidean and Riemannian geometrical approaches \\
	to non--extensive thermo--statistical mechanics \\
%	I. Basic physical and geometrical ideas
}
\author{R. Trasarti-Battistoni}
\address{I.N.F.M. \& Dipartimento di Fisica ``G. Occhialini''\\
Universit\`a degli Studi di Milano-Bicocca\\
Piazza delle Scienze, I-20126 Milano, Italy\\
e-mail: rtb@mib.infn.it}
\date{\today}
\maketitle
%--------------------------------------------------------------------------
\begin{abstract}

We introduce a new geometrical approach to thermo--statistical mechanics.
Here we highlight the main physical ideas,
and how do they translate into geometrical language.
We contrast our approach
with previous (pseudo-)Riemannian [Weinhold 1975; Ruppeiner 1979]
as well as Euclidean [Gilmore 1984; Gross \& Votyakov 1999]
or Euclidean--looking [Levine 1986]
thermo--statistical--geometrical formalisms.
We point out the relevance of our approach within the contexts of 
non--extensive statistical thermodynamics [Tsallis 1988, 2000].
We show how the language of Riemannian geometry
can be used as a powerful investigation tool,
leading to several new and profound questions,
and some temptative answers,
on thermo--statistical mechanics.
\\
{\bf DRAFT} - to be submitted to Phys. Rev. E on XXX 00, 0000; 
accepted                                  on YYY 00, 0000.
\end{abstract}
%\narrowtext
\begin{multicols}{2}
%\twocolumn
%--------------------------------------------------------------------------
\section{Introduction}
\label{Intro}

Geometrical formalisms have been applied to 
statistical thermodynamics in many occasions in the past,
starting with Gibbs himself,
and have regularly prooved themselves to be very useful and insightful
(\cite{Gibbs}; \cite{Tisza}).
In the 1970's and 1980's several other geometrical approaches were proposed
as recently reviewed by Ruppeiner 
(\cite{RuppeinerRev95}; for another geometrical approach see \cite{Rugh97}).
A common feature of all such approaches, 
is the central r$\hat{o}$le played by entropy,
in either of its many forms, namely
the Clausius entropy $S_C$,
the Boltzmann--Shannon entropy $S_B$,
the Gibbs--Shannon entropy $S_G$,
the Boltzmann--Einstein entropy $S_\Omega$.
[See the Appendix for the precise definitions.] 
%(\ref{AppEntro}) 
%
In fact, most of the above-mentioned geometric formalisms
{\it assume} as their start one or the other form of the entropy,
and then {\it derive} from it
their geometric structure, which turns out to be (pseudo-)Riemannian.
[For a deep and very pedagogical treatment of Riemannian geometry,
see \cite{MTW83} or \cite{DNF87}.]
%[See Appendix B.]
%(\ref{AppGeom}).]

However, 
once we are given a well--defined physical system,
{\it what is the ``correct'' entropy we should use?}
Answering such a question is not at all easy
(and the question itself is ambiguous - see \cite{Jaynes57},\cite{Jaynes62}).
Indeed,
following the breakthrough paper of Tsallis (\cite{Tsallis88}),
all along the 1990's an increasing deal of attention has been devoted
to several generalized non--extensive statistical--thermodynamic formalisms
(\cite{TsallisRev00}).
Almost invariably, 
all such investigations postulate as their starting point 
a mathematical form of the entropy functional 
{\it different} from either of the above mentionsd $S$'s,
typically 
the Renyi entropy $S_R$
or
the Havrda--Charvat--Daroczy--Tsallis 
(in brief and nowaday usually, Tsallis') entropy $S_T$
(\cite{HC67},\cite{Daroczy70},\cite{Tsallis88}; see also \cite{Behar}).
% [Appendix (\ref{AppEntro}]
%$S_T({\bf p}) = - \sum_c p^c \log_q p^c$
%where 
%$\log_q x = (1-q)^{-1} (1-x^{-q}) \to \ln x$ for $q \to 1$
%(\cite{Tsallis88}; here $q$ is an exponent, $c$ a label,
%and ${\bf p}$ is a probability distribution over the cells
%in the relevant phase-space).
%
Moreover,
several of the entropy--extremization procedures proposed in such formalisms 
make a different use of the macroscopic constraints $N$, $E$, $V$, and so on
(\cite{Tsallis88};\cite{CT9192};\cite{TMP98}; \cite{MNPP00};
see also \cite{Raggio99}, \cite{dSMOPP98}).
Finally, 
non--extensive thermodynamics 
naturally puts the accent on the r$\hat{o}$le played by 
the total number $N$ of microscopic constituents of the macroscopic system
(though $N$ is usually not explicitly included in the definition of $S_T$
-- but see \cite{JKT95};
\cite{Grigera96};\cite{CT96};\cite{AT98};\cite{CT99}),
and on the effects due to (long--range, or long--time memory)
interactions/correlations among such constituents.

The chief aim of this paper is to provide a convenient framework
able to combine geometry and non-extensivity 
-- and to provide more insight -- into thermo-statistics.
The purpose is of methodological as well as pedagogical nature.
So the obvious question now would be:
can we still derive a useful thermo-stat-geometric formalism
for non-extensive systems?
But since now we have to deal with a modified definition of entropy,
such a question actually splits in two:
(A) can we just take the a new mathematical expression for the entropy
and then derive geometry following the same conceptual path
as for the standard (=extensive) statistical thermodynamics?
Or maybe 
(B) shouldn't we modify as well our approach to combine 
the geometric and thermo-statistics aspects of the problem,
when dealing with non--extensive systems?
We stress that here we do not follow suggestion (A),
which would presumably lead us to some modified 
but still intrinsically Riemannian geometry of thermodynamics.
However, such a program would preliminary require us to settle two key issues:
(A1) is Tsallis entropy the appropriate non--extensive generalization 
of the standard extensive entropy? (And, is it unique?)
Yet, in case it is, (A2) of which of the many entropy forms 
is Tsallis' entropy the generalization? 
(In particular, of $S_B$ or of $S_G$? 
And what about $N$, then?)
Clearly, all such questions have little to do with the specifically
geometric aspect of the problem, and deserve a separate investigation.
So in this paper we rather take on the other road (B). 
From such a point of view,
geometry comes in prior to entropy, which still plays 
a relevant but now secondary r$\hat{o}$le on the scene.
As a consequence, 
we end up with a fundamentally Euclidean geometry,
in which {\it induced} Riemannian structures are present only as far as 
the geometry of the environment space
induces them onto the (curved) embedded submanifolds.
On the other hand, 
the present approach gives a firmer physical handle on such 
(often somewhat confusing) matters as interactions, 
their range, correlations, information, probability, entropy, equilibrium, 
and their multiple conceptual interconnections.

It must be stressed that approaches quite close to ours in spirit
have beeen previously championed by Gilmore (\cite{Gilmore84}),
who lead the Euclidean side in a dispute with the Riemannian 
point of view in the 1980's 
(\cite{Ruppeiner85},\cite{Horn85},\cite{Gilmore85};\cite{ABGIS88}),
and more recently by Gross and Votyakov 
(\cite{GV99},\cite{Gross00}).
In particular,
our formalism turns out to be closely related to those used
by (\cite{SNB85}), (\cite{Levine86}), and (\cite{MNSS90}).
What is new in our approach,
are (i) the attempt to put everything 
in a kinetic--statistical---mechanical (micro/mesoscopic) perspective, 
rather then in a thermodynamical (macroscopic) 
or an information--theoretic ones,
and (ii) an overall constructive attitude, 
i.e. we will always try to define new things
in terms of previously defined objects, 
avoidind assumptions and postulates as much as we can, 
trying hard to find physical justifications otherwise,
and always keeping an explicit and sharp distinction
between the geometry--mathematics and the physics.
Our starting point is classical dynamics of point particles.
Upon that, 
we implement and justify a coarse--graining of 1--particle phase--space.
We then progressively introduce geometrical (Euclidean as well as Riemannian) 
concepts and give their physical interpretation.
(For closely related ideas in different contexts,
see \cite{Dominguez99} and \cite{Kandrup90}.)

The rest of the paper is as follows.
We start in Sect.(\ref{MicroMeso})
with a description of the system we have in mind 
with its microscopic and mesoscopic states, 
and proceed with the macroscopic observables in Sect.(\ref{Macro}).
In Sect.(\ref{Entro}) we translate in our geometrical language
the fundamental thermostatistical concepts of equilibrium and entropy.
In Sect.(\ref{Discu}) we contrast our 
geometric formalism to previous work, and  
discuss the relevance and connections of our work
to non--extensive statistical thermodynamics.
In Sect.(\ref{Conclu}) we conclude.

%--------------------------------------------------------------------------
%%%%%%%%%%%%%%%%%%%%%%%%%%%%%%%%%%%%%%%%%%%%%%%%%%%%%%%%%%%%%%%%%%%%
\section{Micro- and mesoscopic descriptions}
\label{MicroMeso}

Here we take 
the microscopic point of view of kinetic theory, 
and the closely related -- in our opinion, ``mesoscopic'' -- 
point of view of statistical mechanics.
(We will later clarify what we mean here by mesoscopic.)

For the sake of concreteness, the physical system we will consider
in this paper will always be a non--ideal gas of interacting particles,
a paradigmatic example of systems dealt with in standard extensive
statistical thermodynamics, and in its geometrical formulations.
Having always non--extensivity in the back of our mind,
we (i) allow interactions to be long-range, and
(ii) do not deal with any kind of probability 
of being in some more or less likely state, 
but rather with the actually observed 
occupation number of particles within a cell.

The first key idea of our approach is 
to take the $\mu$--space of the gas, 
and divide it into $M$ coarse--grained cells.
At this stage, the partition of phase--space might be taken arbitrarily
-- cells of different shape, size, location, orientation,...
(we might even decide to ``lump'' together in the same cell
portions of phase--space not topologically connected to each other!)
There is only one cell $c(i)$ associated to the $i$-th particle,
but there can be $0\leq n^c \leq N$ particles in the same $c$-th cell.
We will describe the ``coarse--grained microstate'' of the system through
the $M$--tuple ${\bf n}=(n^1,n^2,...n^M)$ of occupation numbers $n^c$'s,
ordered and labeled by $c=1,2,...M$.
We must stress that a state defined by ${\bf n}$ is neither truly microscopic
nor truly elementary, as it ``knows'' little concerning
the arrangement of particles within the cell.
On the other hand, a state described by {\bf n}
is neither truly macroscopic, 
as there are still many ways to shift particles from one cell to another,
or to swap them between two cells,
yet without affecting such macroscopic quantities as the 
total number of particles $N$,
total energy $E$,
total angular momentum $L$, and so on.
To capture the essence of this intermediate status
of {\bf n}, we will therefore call {\it mesostate} from now onward.

The second key idea of our approach is 
to regard the set $N^M$ of all mesostates as a new kind of space
-- ``cell space'' hereafter -- different from both
the $\Gamma$--space and the $\mu$--space.
Geometrically, we will regard the $n^c$'s
as a set of cartesian coordinates in cell space,
and ${\bf n}$ as a vector therein.
We will develop all this in the next Sections.

\subsection{The system: an interacting non--ideal gas}
\label{MicroSystem}

Consider 
a macroscopic system consisting of $i=1,...N$ microscopic constituents.
%each of whom may be found, with a varying degree of probability,
%in any of many ``elementary'' states $c=1,...M$.
%
For the sake of defineteness,
consider a (non--ideal) gas 
of $N$ identical point--like particles 
of mass $m$, position ${\vec r}_i$, and velocity ${\vec v}_i$,
inside a box of volume $V$.
In what follows, the volume will be 
treated as a fixed, external parameter,
independent on the state of the system.

Let each particle in the box 
be immersed in a fixed, external, 
potential field $\phi^{(ext)}({\vec r}_i, {\vec v}_i)$
-- say, 
due to an external electric field ${\vec E}$ or magnetic field ${\vec B}$,
if the particles have electric charge or magnetic moment.
In fact, 
the box itself can be described as a confinining potential well
of the form 
$\phi^{(box)}({\vec r}_i, {\vec v}_i)=0$        for ${\vec r}_i \in V$, 
and 
$\phi^{(box)}({\vec r}_i, {\vec v}_i)=+ \infty$ otherwise.
Finally, let the particles interact with each other 
through a pairwise interaction potential 
$\psi_{ij}=\psi^{(int)}({\vec r}_i,{\vec r}_j;{\vec v}_i,{\vec v}_j)$, 
which in most real cases will just boil down to a simple
position--dependent, pairwise, isotropic, spatially homogeneous potential 
$\psi_{ij}=\psi(r_{ij})$, where $r_{ij}=|{\vec r}_i - {\vec r}_j|$ 
-- say, Van der Waals or Lennard--Jones intermolecular potential, 
or electrostatic Coulomb, or Newtonian gravity.
If we like, we could also take $\psi(0)=0$, 
to avoid embarassing self--energies.
This is but a formal way of escaping from physical paradoxes
raised by the mathematical idealization of point-like particles. 
More rigorously, one should take into account a finite, non-zero
size of the particles themselves.

On the microscopic, kinetic--theory level, 
the state of the system would be completely specified if we knew 
all 
the $6N$ coordinates in the $N$--particle phase--space ($\Gamma$--space).
A theoretically (much) less informative, 
but experimentally (a lot) more accessible description is via
the $6$--dimensional $1$--particle phase--space ($\mu$--space) of the gas,
where inter--particle correlations are either totally ignored,
or retained just at the mean--field level of approximation.
Ideally, 
both descriptions are as fine as possible:
a microstate is an $N$--dimensional point 
$\delta^N_D(\vec x_1(t),...\vec x_N(t)
;\vec v_1(t),...\vec v_N(t))$ in $\Gamma$--space, or
a collection of $N$ (possibly interacting) $6$--dimensional points 
$\delta^6_D(\vec x(t); \vec v(t))$ in $\mu$--space.

So far, we made no use of any ``elementary'' state $c=1,...M$;
at this stage, our states are a lot more than just $M$, in fact 
they are $\infty^{6N}$ in $\Gamma$--space and $\infty^{6}$ in $\mu$--space.
Neither did we make any ``statistical'' assumption so far --
we only borrowed concepts from classical mechanics.
Only now will we move on to cells and ``mesostates'' ${\bf n}$.
This will allow us to introduce a new arena -- ``cell space'' -- 
inside which geometry will later come to perfom.
Prior to that, however, we will need to introduce
and physically justify a few statistical--dynamical assumptions.

%%%%%%%%%%%
\subsection{Cell space: coarse--graining,
information, ``effective mean--field'', partition, covariance}
\label{MicroCellSpace}

First of all, 
why coarse--graining?
Physically, it is very well motivated.
In real experiments or even in numerical simulations
we have access only to a coarse--grained kind of information,
since 
(i) real--life results are always affected by 
some non--zero uncertainty and/or numerical cut--offs, 
which give an effective coarse--graining, and
(ii) anyway we often group experimental results into bins after all.
But then, 
why $\mu$-space?
Well, again, in real--life the information we have most easily access to is
typically that contained within the $1$--particle description,
rather than that of the full $N$--particle description.
Moreover, the kind of physical understanding we have/seek of a system
usually involves just 1- or 2-particle objects, 
like kinetic or pairwise interaction energy, respectively;
very rarely do we need to consider any intrinsically 
$3$-, $4$-, or $N$-body quantity.
We then end up with a new statistical description,
where the relevant coordinates are now
the cell occupation numbers $n^c=n^1,n^2,...n^M$, 
a state is identified by ${\bf n}$, and 
the relevant ``phase''-space will be from now on called cell--space,
and sometimes denoted as $N^M$.

From the point of view of information theory,
going from $\Gamma$--space to $\mu$--space
makes a less efficient use of the information at our disposal,
insofar as it ignores correlations among particles.
Going from $\mu$--space to cell--space
introduces a further loss of information,
since all particles within the same $c$--th cell are now equivalent, 
while they are still different in $\mu$--space.
So why should we deliberately choose to loose so much information,
first taking $\mu$--space and then even coarse--graining it?
Well, we might be so clever and so lucky to be able to 
throw away {\it most} of the information we might keep
but would not need, and keep {\it just} that we are really interested in.  
This is exactly the basic philosophy behind the use of
expectation values of macroscopic variables as constraint
(``available information'') on the search for the maximum--likelihood 
probability distribution, i.e. the standard practice in the
information theory approach to statistical thermodynamics (\cite{Jaynes57}).
However, 
after the obvious macroscopic constraints,
we might -- and should! -- try to include 
in our description whatever useful extra information we have on the system.
So we should also include any available microscopic information, if useful.

It should not be overlooked, however, that 
coarse--graining introduces several physical approximations in what follows.
As an example, let us focus only on 
position--dependent, pairwise, isotropic, homogeneous
interactions $\psi(r_{ij})$, for simplicity. 
(Similar considerations should apply as well to more general cases, 
even velocity--dependent, interactions.)
The {\it particle}'s position and velocity ${\vec r}_i$, ${\vec v}_i$
will only approximately coincide with the {\it cell}'s position and velocity
${\vec r}_c \approx {\vec r}_i$, ${\vec v}_c \approx {\vec v}_i$.
As a first step, 
we could just assume this approximation to be good for our purposes. 
But then, 
two particles within the same cell should be regarded as separated 
by a distance $r_{ij}\approx 0$, as long as $r_{ij}\leq l_c$, 
where $l_c$ is the typical linear size of the $c$-th cell.  
As a consequence, 
we would be tempted to approximate their mutual interaction energy
per pair of particles by 
$\frac{1}{2} \psi_{cc} \approx \frac{1}{2} \psi(0)$,
which will either 
(a) usually lead to mathematical divergences and unphysical paradoxes
if we take the true $\psi(r)$ evaluated at $r=0$, 
or 
(b) be a bad approximation within the cells if we force $\psi(0)=0$.
Of course, 
two rough solutions could be
(c) just take 
$\psi_{cc} \approx \psi(l_c)$,
or better 
(d) assume a uniform density 
$n(\vec r \approx \vec r_c) \approx n^c/l_c^3$ 
inside the $c$-th cell and compute the resulting $\psi_{cc}$
by spatial integration of $\psi(r_{ij})$ over $ij$-pairs in the cell.
As a second step, we might try to improve on such a situation as follows.
Though we do not exactly how and why, 
we may {\it expect} particles within a given cell, 
as a consequence of their dynamical evolution,
to arrange themselves in some well--defined, typical way, 
depending only on the 
(size, shape, location, orientation, and possibly occupation of the) cell.
As a concrete example, 
it is well-known that many self-gravitating systems
-- star clusters, galaxies, galaxy clusters -- 
tend to assume spherical configurations
with a few well--defined and to some extent ``universal'' profiles 
(not simply uniform!) of density, velocity, potential, and so on.
Explicitly, we make two assumtpions:
(1) dynamically, particles arrange their positions
in some well--defined way around any given point ${\vec r_*}$ in space,
such that there exist a ``typical'' (say, ``mean'') density profile 
$
n({\vec r}) \approx {\overline n} f(|\vec r_* - \vec r|) 
$
where ${\overline n}=N/V$ is the global mean number density;
(2) statistically, such a profile is ``typical enough'',
i.e. it does not vary (too much) from one (occupied) point to another.
As central point ${\vec r_*}$ we may then take as well 
the cell center ${\vec r_c}$.
We may now integrate over the cell 
(hence $r_i=|\vec r_i - \vec r_c|$) 
and obtain:
\begin{equation}
\label{nclc}
n_c(l_c) \approx
F_c(l_c) :=
{\overline n} 
%\int_{0\leq |\vec r_i| \leq l_c} d^3{\vec r_i} 
\int d^3{\vec r_i}
f(r_i)
\end{equation}
and 
\begin{equation}
\label{psicclc}
\psi_{cc}(l_c) \approx
G_c(l_c) :=
{\overline n}^2
%\int_{0\leq |\vec r_i| \leq l_c}  d^3{\vec r_i}   
%\int_{0\leq |\vec r_j| \leq l_c} d^3{\vec r_j}  
%f(r_i) f(r_j) \psi(|\vec r_i - \vec r_j|)
\int d^3{\vec r_i}
\int d^3{\vec r_j}  
f(r_i) f(r_j) \psi_{ij}
\end{equation}
Note that (\ref{nclc})
can be viewn as as a local mean (=over the cell) density, 
but not as the uniform global mean (=over the whole volume) density. 
Through $f(r)$, it still retains some (=the ``typical'') informations
regarding the local arrangements in space 
of the particles in the neighbourhood of $\vec r_c$.
On the other hand, once the typical profile $f(r)$ is given, 
(\ref{psicclc}) is immediately recognizable as the
mean--field estimate of the 2-particle potential,
ignoring the correlations among particle pairs within the cell, 
i.e. correlated fluctuations leading to 
departures of the actual density distribution from the typical $f(r)$.
Now we assume $n_c(l_c)$ is invertible
(which is physically reasonable given a monotonic $\psi({\vec r})$),
we invert (\ref{nclc}) for $l_c$, substitute into (\ref{psicclc}) 
to eliminate $l_c$ in favour of $n_c$, and obtain:
\begin{equation}
\label{psiccnc}
\epsilon_c^{\psi}(n^c) \approx \psi_{cc} (l_c(n_c))
\end{equation}
which appears as an effective cell--cell interaction potential
when ``seen from the outside'' of the $c$--th cell.
It depends on the occupation ${\bf n}$
through the cell occupation $n^c$, 
but it depends on properties {\it of the $c$-th cell only}, 
memory-less of what happens to the particles inside
and/or to the particles in other cells.
Note that 
near--neighbour--cell interactions are on scales  $r\geq l_c$, 
i.e. still not too far from 
the intra--cell interactions on scales $r\leq l_c$
we just considered at step two. 
As a third step, we could then hope and try to model also
such a situation in some cell--dependent, but particle--independent, way.
So, presumably, (\ref{nclc}) ,(\ref{psicclc}), and (\ref{psiccnc})
would be generalized to
$n_c \approx F_c(l_c;{\bf n})$,
$\psi_{cc} \approx G_c(l_c;{\bf n})$,
and $\epsilon_c^{\psi}(n^c;{\bf n})$, respectively,
now depending on properties all the $M$ cells,
but still just on {\it cell} properties. 

In other terms, when we decide to {\it ignore completely} 
how particles are arranged within the cells in phase--space,
and consider only the intra--cell interactions,
we are making a mean--field approximation,
supposing that correlations among the particle's
positions and velocities within the cells
(and between near--neighbour cells) are totally neglectable.
In this mean--field description, we only deal with
cells and properties of cells.
However, we may as well decide to {\it ignore only partially} 
the internal arrangements of particles within the cells,
while {\it retaining some} information 
(that which is most relevant to our purposes) 
concerning particles interactions/correlations.
In other words, if we find it useful, 
we may consider particles in the same cell 
to be at a $r=0$ distance from each other 
as far as we are concerned their position $\vec r \approx \vec r_c$ 
but at the same time remind that they are at arranged in some prescribed way
with non-zero distances as long as we are interested in
their interaction $0 \not = \psi_{cc} \not = \infty$.
The ``typical'' cell--dependent internal arrangements
should be regarded exactly as that extra bit of useful information
we alluded to hereabove.
Now, if the scenario summarized in 
(\ref{nclc}) ,(\ref{psicclc}), and (\ref{psiccnc}) were actually viable,
we would then end up with solution (e): 
an ``effective'', `` mean--field--like'' description, 
quantitatively different from the pure mean--field,
but qualitatively quite close to it.
Whether and how exactly such a program can be concretely implemented 
(as the results of \cite{Dominguez99} seem to imply)
will of course depend on the specific kind of interactions 
(and the consequent internal arrangements) we must deal with, 
which would take us too far off the main path of this paper. 
Here we will just content ourselves with assuming that
such a mean--field--like description is actually feasible,
and to investigate the geometrical and thermo--statistical consequences
of such a statistical--dynamical assumption.

Finally, 
{\it the way we choose} the partition of phase--space is fundamental.
Before deciding whether to include little/some/all of
the available informations concerning {\it particles} in cells,
we must still fully specify how the $M$ {\it cells} themselves
look like (size, shape, location, orientation),
independently on their occupation number $n^c$.
Such a  choice might a clever, or a not--so--clever one.
In order to be useful, it should be guided by physical intuition,
based on our overall knowledge of the physics of the system. 
Our geometrical--thermo--statistical formalism cannot help us anyhow here.
Viceversa, it will be completely determined by 
(i) our choice of the ``foremost'' coarse--graining, and
(ii) the ``background'' physics of the system
(be it dominated by intra-cell, near(est)--neighbour-cell, or inter--cell
interactions/correlations).
This very same point has been stressed by Jaynes himself
(\cite{Jaynes57}), 
in his remarks upon the ``antropomorphic'' nature of entropy; 
the only difference with us is that
he restricted himself to $Q$=a few macroscopic thermodynamical coordinates
$N,E,V,...$,
while we extend the discussion to $M$=many mesoscopic coordinates 
$n^1,...,n^c,...n^M$.
The choice of a partition can equivalently be interpreted
as a choice of {\it how to weight} (as opposed to simply {\it how to count}) 
the available mesostates; 
a physically well--motivated choice would be weighting a state
by the time the system actually spends therein
(\cite{SB87}).
In any case, 
as here we need not rely on any specific choice 
neither of the partition nor of the entropy,
we are free to ignore such difficulties now and to leave them
to later investigations.

%%%%%%%%%%%
%Figure 1? gamma -> mu -> cell
%%%%%%%%%%%

%%%%%%%%%%%
{\subsection{Mesoscopic states: cartesian \\
coordinates and vectors in cell space}}
\label{MicroVec}

What is the geometrical meaning of a mesostate ${\bf n}$?
By construction, cell space is equipped 
with a set of coordinates $n^c$'s,
which we from now on will regard as cartesian and orthogonal.
There is a set of preferred directions along which 
only one coordinate grows, but the other stay constant 
-- i.e., a set of unit vectors ${\bf e_1},...{\bf e_M}$ 
associated with the coordinate system $n^1,...n^M$ --
and a set of preferred hyperplanes $\Sigma_{n^c}$,
upon which
only the $c$-th coordinate stays constant, while all the others may change.
For clarity,
in what follows repeated index will not be implicitly summed over,
but we will always provide the explicit summation symbol 
(not a surface!) $\sum_{c=1}^M$.

Each mesostate of the system can be regarded as
an $M$--dimensional vector (components with upper indexes):
\begin{equation}
\label{n}
{\bf n}
	:=         \sum_{c=1}^M n^c {\bf e}_c
\end{equation}
The unit vectors themselves have components
$({\bf e}_a)^b = \delta_a^b$
(where $\delta_a^b=1$ iff $a=b$, and $\delta_a^b=0$ otherwise),
i.e. 
${\bf e}_c$ is the mesostate ``only 1 particle, in the $c$-th cell''.
Physically, ``moving along the direction of ${\bf e}_c$'' 
means simply to add/remove a particle from the $c$-th cell,
leaving any other $h$-th cell untouched,
i.e.
$({\bf e}_c)^b = \frac{\partial n^b}{\partial n^c}=\delta^b_c$.

Let us define the vector sum between two mesostates
in the most natural way, 
by summing their components:
\begin{equation}
\label{n+n}
({\bf n}' + {\bf n}'')
	:=         \sum_{c=1}^M \left[(n^c)'+ (n^c)''\right]{\bf e}_c
\end{equation}
and the product of a mesostate vector with a scalar $x$:
\begin{equation}
\label{xn}
x\cdot{\bf n}
	:=         \sum_{c=1}^M (x \cdot n^c) {\bf e}_c
\end{equation}
Note that the definitions (\ref{n+n}) and (\ref{xn}) 
were not introduced as related to any
``physical operation'' (compressing/expanding, 
adding/removing energy, merging/splitting, etc.)
effectively involving one or two different physical systems.
Mathematically, we will soon need them in Sect.(\ref{MicroDirLenDist}).
%We will later meet other physical reasons
%justifying the introduction of (\ref{xn}) in Sect.(\ref{EntroEquiTraj}).

%%%%%%%%%%%
\subsection{Diversity and similarity of mesostates: 
(euclidean) distance, direction, and length}
\label{MicroDirLenDist}

Physically speaking, we would like to regard 
two mesostates ${\bf n}'$ and ${\bf n}''$ as being
``close to each other'' if their respective occupation numbers
are as similar as possible.
This naturally leads to a notion of (euclidean) distance:
\begin{equation}
\label{dnn}
d({\bf n}',{\bf n}'')
	:= \sqrt{\sum_{c=1}^M |(n^c)''- (n^c)'|^2}
\end{equation}
telling us ``how much'' ${\bf n}'\neq {\bf n}''$.
Two mesostates coincide iff $d({\bf n}',{\bf n}'') = 0$.
However, 
$d({\bf n}_0,{\bf n}') = d({\bf n}_0,{\bf n}'')0$
does not necessarily imply ${\bf n}'={\bf n}''$;
there can be many mesostates upon the $M-1$--dimensional sphere
of radius $d$ and centre ${\bf n}_0$,
all at the same distance $d({\bf n}_0,{\bf n}) = d$.

It would be useful to have some other mathematical tool
telling us ``in which way'' ${\bf n}'\neq {\bf n}''$.
For example, all unit vectors ${\bf e}_c$'s 
differ all by the same extent (=same distance $d=1$) from 
the ``completely empty mesostate'' ${\bf 0}$
(no particles anywhere at all),
but they are completely different mesostates;
we would like to describe such a state of affairs quantitatively.
It is also natural to regard the ${\bf e}_c$'s 
as being orthogonal to each other,
${\bf e}_c \parallel {\bf e}_c$ and 
${\bf e}_a \perp {\bf e}_b$ (for $a \neq b$). 
Such directional ideas can be made more precise as follows.
First, we define a scalar product 
between any two mesostates ${\bf n}'$ and ${\bf n}''$ 
by orderly multiplying their components, and summing them up:
\begin{equation}
\label{nn}
{\bf n}'\cdot {\bf n}''
	:= \sum_{c=1}^M (n^c)' \cdot (n^c)''
\end{equation}
Then we define the modulus -- ``length'' -- of a mesostate:
\begin{equation}
\label{|n|}
|{\bf n}|
	:=\sqrt{ \sum_{c=1}^M (n^c)^2  }
	=\sqrt{ {\bf n} \cdot {\bf n} }
\end{equation}
which implies $|{\bf n}|\leq N$ (equal iff all particles are in the same cell).
We can now formally introduce the concept of direction in cell space
through the unit vector ${\bf u}_n={\bf n}/{|{\bf n}|}$:
a direction is the equivalence class of all mesostates
sharing the same ${\bf u}_n$.
(Note that the direction vector ${\bf u}_n$ and the probability vector
${\bf p}= {\bf n}/N = {\bf u}_n \cdot {\bf |n|}/N$ 
have different normalizations.)
Finally, we define
${\bf n}' \perp {\bf n}''$ iff ${\bf u}_n' \cdot {\bf u}_n''=0$
and 
${\bf n}' \parallel {\bf n}''$ iff ${\bf u}_n' \cdot {\bf u}_n''= 1$.
This allows us to clarify the notion of ``orthogonal unit vectors'' 
implicitly used in (\ref{n}):
for each $c$, $|{\bf e}_c| = \sqrt{ {\bf e}_c \cdot {\bf e}_c }=1$,
and for each $a \not = b$,  ${\bf e}_a \cdot {\bf e}_b =0$.

There is an easy physical interpretation of (\ref{nn}) and (\ref{|n|}).
Note that, since $0\leq (n^c)',(n^c)''\leq N$, by definition
occupation numbers can never be negative, and
$(n^c)' \cdot (n^c)'' \geq 0$ for each $c$ in (\ref{nn}).
So if ${\bf n}' \cdot {\bf n}''=0$ then 
for any $c$, either $(n^c)'=0$, or $(n^c)''=0$, or both.
That is, ${\bf n}' \perp {\bf n}''$ actually means
there are no directions along which 
${\bf n}'$ and ${\bf n}''$ have simultaneously non--zero components
-- two orthogonal mesostates are totally different,  
having no cell at all in common.
On the other hand, ${\bf n}' \parallel {\bf n}''$ means 
two parallel mesostates ``occupy'' exactly the same cells, 
with a possibly different occupation  ${\bf n} = {\bf u}_n|{\bf n}|$,
and each component $n^c \propto N$.
If ${\bf n}'$ and ${\bf n}''$ coincide both in modulus and direction, 
they are just the same mesostate.
What if ${\bf n}' \neq {\bf n}''$, 
but ${\bf n}' \cdot {\bf n}'' \approx |{\bf n}'| |{\bf n}''|$?
The natural interpretation is that the mesostates 
${\bf n}'$ and ${\bf n}''$ are somehow ``similar'',
and we expect $d({\bf n}',{\bf n}'')$ to be in some sense small.

So, 
while (\ref{dnn}) 
gives a quantitative measure of the ``diversity'' of mesostates,
(\ref{nn}) gives a quantitative measure of their ``similarity''.
No surprise then, that (\ref{dnn}), (\ref{nn}), and (\ref{|n|})
are strictly intertwined with each other.
In fact, 
setting $\Delta {\bf n}={\bf n}''-{\bf n}'$,
we can recast (\ref{dnn}) as 
$d({\bf n}'',{\bf n}')
=\sqrt{ \Delta {\bf n} \cdot \Delta {\bf n}}
=|\Delta {\bf n}|$ --
another motivation to introduce (\ref{n+n}) and (\ref{|n|}).
Viceversa, 
$|{\bf n}|
= \sqrt{ {\bf n} \cdot {\bf n}}
=d({\bf n},{\bf 0})$.
Also, 
we might recast (\ref{nn}) in terms of (\ref{|n|}):
\begin{equation}
\label{nn|n|}
{\bf n}' \cdot {\bf n}''
	= \frac{1}{2}
	\left[ |{\bf n}'+{\bf n}''|^2-(|{\bf n}'|^2+|{\bf n}''|^2)\right]
\end{equation}
or also:
\begin{equation}
\label{nn|n|bis}
{\bf n}' \cdot {\bf n}''
	= \frac{1}{4}(|{\bf n}'+{\bf n}''|^2-|{\bf n}'-{\bf n}''|^2)
\end{equation}
as can ben checked by direct substitution.
Equation (\ref{nn|n|bis}) shows explicitly that
the projection coefficient ${\bf n}' \cdot {\bf n}''$
tells us ``how much'' ${\bf n}' \approx {\bf n}''$, so to say.
In fact, 
when ${\bf n}' = {\bf n}''$ the second term in (\ref{nn|n|bis}) vanishes
and we get the maximum value
$({\bf n}'\cdot {\bf n}'')_{max} = +|{\bf n}'||{\bf n}''|$,
while 
when ${\bf n}' = - {\bf n}''$ the first term in (\ref{nn|n|bis}) vanishes
and we get the minimum value
$({\bf n}'\cdot {\bf n}'')_{min} = -|{\bf n}'||{\bf n}''|$.
Analogous arguments also prove that 
${\bf n}' \cdot {\bf n}'' \approx |{\bf n}'| |{\bf n}''|$
is equivalent to $d({\bf n}',{\bf n}'') = |\Delta {\bf n}| \approx 0$, 
as expected.
Finally, 
we have all the usual properties of euclidean distance and scalar product, 
i.e. Pythagora's and Euclide's theorems (\ref{nn|n|})-(\ref{nn|n|bis}),  
the Schwar(t?)z inequality:
\begin{equation}
\label{schwar(t?)z}
{\bf n}' \cdot {\bf n}''
	\leq |{\bf n}'||{\bf n}''|
\end{equation}
and the triangular inequalities:
\begin{equation}
\label{tri+}
|{\bf n}' + {\bf n}''|
	\leq |{\bf n}'|+|{\bf n}''|
	=   ||{\bf n}'|+|{\bf n}''||
\end{equation}
\begin{equation}
\label{tri-}
|{\bf n}' - {\bf n}''| 
\geq
||{\bf n}'|-|{\bf n}''||
\geq 
|{\bf n}'|-|{\bf n}''|
\end{equation}
As far as we are aware, 
the notion of an {\it euclidean} distance among thermodynamical states
has been previously emploid only by (\cite{Levine86}).
There are several differences with our approach, 
which we discuss in Sect.(\ref{Discu}).

%%%%%%%%%%%
\subsection{Euclidean metric - why? \\
%(metric and fluctuations)
}
\label{MicroWhyEuclidean}

From a mathematical point of view, 
we have just seen several good reason to choose 
(\ref{dnn}) as our definition of distance.
Say, we might instead have raised the absolute value in (\ref{dnn}) 
to some other exponent -- 1 or 4 or 15 etc. -- 
and we would have still found a quantitative way to describe
mesostates as different.
However, such other distances would have not been euclidean,
i.e. Pythagoras' theorem etc. would not hold.
And we wouldn't have been able to introduce a scalar product 
like in (\ref{dnn}), and to use it to quantify the idea of
similarity of two states as in (\ref{nn}).
In general,
we would have denied ourselves the use of a whole arsenal
of well--known mathematical tricks and results of euclidean geometry,
like (\ref{nn})--(\ref{tri-}),
without gaining much as a quantitative notion of distance.

Geometrically speaking, the {\it euclidean distance} (\ref{dnn})
endows the cell space $N^M$ with a {\it flat, euclidean metric}.
(Technically, one says the pair $(N^M,\delta_{ab})$
is an euclidean manifold.)
Defining the (finite) displacement $\Delta n^c := (n^c)' - (n^c)''$, we have:
\begin{equation}
\label{dndeltan}
d^2({\bf n}',{\bf n}'')
	= \sum_{a=1}^M \sum_{b=1}^M \Delta n^a \delta_{ab} \Delta n^b
\end{equation}
where $\delta_{ab}=1$ iff $a=b$, and $\delta_{ab}=0$ otherwise.
Note that 
the euclidean metric tensor $\delta_{ab}$ 
is independent on the position ${\bf n}$, and
this is why we may use the finite expression (\ref{dndeltan});
given a more general ``position''--dependent metric tensor
$g_{ab}({\bf n})$ we should use instead:
\begin{equation}
\label{dnpath}
d^2_g({\bf n}',{\bf n}'')
	= 
\left[ 
\int_{0}^{1} d\lambda 
\sqrt{ 
\left(
\sum_{a=1}^M \sum_{b=1}^M 
\frac{dn^a}{d\lambda} \frac{dn^b}{d\lambda} g_{ab}
\right)
} 
\right]^2
\end{equation}
where $\lambda \in [0,1]$ is a suitable affine parameter that describes
the trajectory ${\bf n}(\lambda)$ between ${\bf n}'$ and ${\bf n}''$, 
and one must further specify
exactly {\it which} trajectory in $N^M$ is being followed --
i.e., the distance between two mesostates would not be unique,
but it would depend on the path along which it is measured.
(Which is reminiscent of equlibrium/non--equilibrium transformations).
One might prefer to refer to (\ref{dnpath}) as ``length'' (of a path)
between ${\bf n}'$ and ${\bf n}''$, 
and to reserve the term ``distance''
to, say, the minimum (=geodesic) path--length  
between ${\bf n}'$ and ${\bf n}''$.

Physically,
a metric tensor which is ``everywhere the same'' in $N^M$,
deals with every mesostate on exactly the same footing.
Of course,
different mesostates will still possess physically different properties,
and some mesostates might be physically preferred in some sense
-- say, they might be equilibrium states,
extremizers of this or that thermodynamic potential, and so on.
However, once a partition as been chosen,
such (real!) differences will raise only due to the underlying physics, 
but they will not be ``artificially'' induced
by the overimposed mathematical structures
(\ref{n})--(\ref{|n|}).

%Physically, there is another good reason to choose 
%(\ref{dnn}) as ``the'' distance in cell space.
%Typically, thermodynamic quantities fluctuate around some mean value,
%and we would correspondingly like to be able to speak about
%the deviations of a mesostate ${\bf n}=\hat {\bf n} + \Delta {\bf n}$ 
%around a given mesostate $\hat {\bf n}$.
%But then%
%
%\begin{equation}
%\label{drms}
%\frac{1}{\sqrt{M}} d(\hat {\bf n},\hat {\bf n}+\Delta {\bf n})
%	=\sqrt{ \frac{1}{M}\sum_{c=1}^M |\Delta n^c|^2 }
%\end{equation}
%
%is simply the r.m.s. deviation
%of the $n^c$'s of ${\bf n}$ around $\hat {\bf n}$,
%which is typically something accessible from experiments.
%More exactly,
%repeating experiments amounts to many realizations of (\ref{drms}),
%i.e. many different mesostates 
%${\bf n}',{\bf n}'',{\bf n}''',...$
%fluctuating around $\hat {\bf n}$
%at a typical distance 
%$d_{r.m.s.}=
%\sqrt{\langle d^2(\hat {\bf n},\hat {\bf n}+\Delta {\bf n})\rangle_M}$
%where $\langle ... \rangle_M$ denotes 
%an ensemble average over $M$ experiments.
%
%As far as we are aware, 
%the connection among a ``thermodynamical metric''
%and the property of fluctuations among thermodynamical mesostates
%has been first pointed out by 
%Ruppeiner (\cite{Ruppeiner79},\cite{RuppeinerRev95}).
%Basically, the differences with our approach are
%(i) a space with a lot fewer directions than cell--space,
%and 
%(ii) a different (entropy--based) definition of distance. 
%We discuss this in Sect.(\ref{Discu}).

%%%%%%%%%%%%%%%%%%%%%%%%%%%%%%%%%%%%%%%%%%%%%%%%%%%%%%%%%%%%%%%%%%%%

\section{Macroscopic description}
\label{Macro}

Here we take the macroscopic point of view of (statistical)
thermodynamics. Typically, we are interested in $Q=$a few 
(and mostly mechanical) macroscopic quantities $A$, 
such as number of particles, pressure, energy, etc.
Here we do not consider their mean (=per particle) value $A/N$
and/or expected (=ensemble averaged) value $\langle A \rangle$,
but 
their total and actually observed (or observable, in principle) value $A$,
consistently with our attention to non--extensive and concrete matters.
Entropy requires separated additional considerations, 
postponed to Sect(\ref{Entro}).

\subsection{Macroscopic constraints \\
and mesoscopic functionals}
\label{MacroSystem}

Consider a physical system,
whose macroscopic thermodynamical state can be
completely characterized by giving a few numbers $A^*$
-- say, total number of particles $N^*$ and total energy $E^*$. 
Moreover, suppose we know how $N^*$ and $E^*$ could be computed
if we only knew the mesoscopic state ${\bf n}^*$ the system is in.
In other words, for each $A$
we are given a corresponding mesoscopic functional $A(...)$, 
whose argument is a mesostate.
Without loss of generality, we can express $A({\bf n})$ as follows:
\begin{equation}
\label{Asum}
A({\bf n})
	:=	\sum_{c=1}^M n^c \cdot a_c({\bf n})
%	= N	\sum_{c=1}^M p^c
%	= N	1
\end{equation}
where the arbitrary dependence of $A$ on ${\bf n}$ has been reparted
between the $a_c$'s and the $n^c$'s for reasons to become clear shortly.
In particular, we are given $N(...)$ and $E(...)$, 
which when evaluated at an arbitrary mesostate ${\bf n}$ yield:
\begin{equation}
\label{Nsum}
N({\bf n})
	=	\sum_{c=1}^M n^c
%	= N	\sum_{c=1}^M p^c
%	= N	1
\end{equation}
and
\begin{equation}
\label{Esum}
E({\bf n})	
	=	\sum_{c=1}^M n^c  \cdot \epsilon_c
%	= N	\sum_{c=1}^M p^c  \cdot \epsilon_c
%	= N	\langle 	 \epsilon_c	\rangle
\end{equation}
while, when evaluated at some mesostate ${\bf n}^*$
compatible with the macroscopic constraints $N^*$ and $E^*$, 
yield $N({\bf n}^*)=N^*$ and $E({\bf n}^*)=E^*$.
The energy levels $\epsilon_c$'s are $M$ known mesoscopic parameters,
possibly themselves depending on ${\bf n}$
and/or on some external parameter $X$
(Sect.\ref{MacroRiemCurv}, \ref{MacroExt/Kin/Int}).
Knowing the mesostate ${\bf n}$ 
automatically fixes the values of $N$ and $E$,
but the reverse is not true; there might be many, 
different states ${\bf n}' \neq {\bf n}'' \neq {\bf n}'''$ etc.
all compatible with the same pair $(N,E)$.

We will assume the functionals $A(...)$
to be continuous and smooth enough as needed,
so that we might treat the discrete coordinates $n^c$'s
as if they were continuous variables, and use
continuity, derivatives, integrals, and so on.
Mathematically, the variations should be small, which means
$|\Delta n^c| << n^c$ for each $c$ and/or $|\Delta{\bf n}| << |{\bf n}|$.
Physically, $N$ and $E$ should change very slowly with ${\bf n}$.
%We might expect this to make sense if 
%$n^c << N$ for each $c$ and/or $|{\bf n}| << N$
%(a distribution function sufficiently spread upon many cells).
%
%[{\bf ??? is $|{\bf n}|$ small or large??? rethink it a while.}]
%
This is then expected to breakdown in proximity of phase transitions
and/or at critical points,
where even a tiny change of the system's state 
can cause enormous modifications to its properties.

%%%%%%%%%%%
\subsection{Macroscopic observables: covectors in cell space}
\label{MacroCoVec}

What is the geometrical meaning of a macroscopic observable $A({\bf n})$?
There is a geometrical interpretation of (\ref{Asum}), similar to (\ref{n}), 
in terms of a suitable ``observable covector'' , 
or covariant $1$--form (\cite{Levine86}).
First of all, let us associate 
the $M$--tuples
${\underline a}=[a_1,a_2,...a_M]$,
${\underline 1}=[1,1,...1]$, and 
${\underline \epsilon}=[\epsilon_1,\epsilon_2,...\epsilon_M]$
to the mesoscopic functionals 
$A(...)$, $N(...)$, $E(...)$, respectively.
Then, 
let us define the observable covector ${\underline A}$
(components with lower indexes) as:
\begin{equation}
\label{A}
{\underline A}
	:=         \sum_{c=1}^M a_c({\bf n}) \cdot {\underline e}^c
\end{equation}
where the unit covectors correspond to 
the mesoscopic functionals with components
$({\underline e}^a)_b = \delta^a_b$,
i.e. 
${e}^c({\bf n}):=n^c$ is that particular functional telling us
how many particles are within the $c$-th cell.
(Implicitly, we have so defined 
the sum ${\underline A}'+{\underline A}''$ of two covectors,
and 
the product $x\cdot {\underline A}$ of a scalar with a covector, 
in a manner totally analogous to (\ref{n+n}) and  (\ref{xn}).)
Finally, let us define the mixed scalar product of 
a mesostate vector ${\bf n}$ and an observable covector ${\underline A}$
as follows:
\begin{equation}
\label{Ascalar}
{\underline A} \cdot {\bf n} 
	=         \sum_{c=1}^M a_c({\bf n}) \cdot n^c = A({\bf n})
\end{equation}
Clearly, 
(\ref{Nsum}) and (\ref{Esum}) may now be rewritten as:
\begin{equation}
\label{Nscalar}
N({\bf n})
	= {\underline 1} \cdot {\bf n} 
\end{equation}
and
\begin{equation}
\label{Escalar}
E({\bf n})	
	= {\underline \epsilon} \cdot {\bf n} 
\end{equation}
As already sketched,
the occupation numbers $n^c$'s themselves
may be viewn as observable covectors;
the associated $M$--tuples are the $M$ $unit 1$--forms 
${\underline e}^c=[0,0,...,1,...0]$
where the $1$ is in the $c$-th position,
and we get:
\begin{equation}
\label{nscalar}
n^c({\bf n})
	= \sum_{b=1}^M \delta^c_b \cdot n^b
	= {\underline e}^c \cdot {\bf n} 
\end{equation}

In the same spirit,
given a physical observables $G({\bf n}',{\bf n}'')$ 
involving 2 mesostates we can define the associated $2$--nd rank 
(co)tensors ${\underline {\underline G}}$;
e.g. the metric tensor ${\underline {\underline \delta}}$ 
corresponding to (\ref{dndeltan}) which can be compactly rewritten as:
\begin{equation}
\label{dndeltanscalar}
d^2({\bf n}',{\bf n}'')
	= {\bf \Delta n}' \cdot 
{\underline {\underline \delta}} \cdot {\bf \Delta n}''
\end{equation}
and analogously for (\ref{dnpath}).

As far as we are aware, the idea of associating 
a macroscopic observable $A$ to a covector ${\underline A}$ 
is  originally due to Levine (\cite{Levine86}).
There are several differences with our approach, 
which we discuss in Sect.(\ref{Discu}).

%%%%%%%%%%%
%Figure 2? cellspace / vectors / surfaces / intersection
%%%%%%%%%%%

%%%%%%%%%%%
\subsection{Macroscopic constraints: scalar fields \\
and level (curved) surfaces}
\label{MacroConstrSurf}

There is another geometrical interpretation of (\ref{Nsum}) and (\ref{Esum}): 
$N({\bf n})$ and $E({\bf n})$ are two scalar fields, 
depending on the ``position'' ${\bf n}$, 
and defined everywhere in cell space.
Fixing specific values $N=N^*$ and $E=E^*$
singles out two $(M-1)$--dimensional ``level surfaces''
$\Sigma_N^* = \Sigma_N(N^*)$ and $\Sigma_E^* = \Sigma_E(E^*)$ within $N^M$.
Any mesostate 
${\bf n^*}'$ upon $\Sigma_N^*$ will yield $N({\bf n^*}')=N^*$, 
and similarly 
if ${\bf n^*}''$ upon $\Sigma_E^*$ then $E({\bf n^*}'')=E^*$.
Constraining ${\bf n^*}$ to be {\it simultaneously} compatible
with $N^*$ and $E^*$ means to considerate only those mesostates
belonging to the $(M-2)$--dimensional intersection
$\Sigma_N^* \cap \Sigma_E^*$, 
i.e. the ``constraint surface'' 
$\Sigma_C^*=\Sigma_C(N^*, E^*)=\Sigma_N(N^*) \cap \Sigma_E(E^*)$.
(See also Sect.(\ref{EntroExtr}).)

In particular, 
since (\ref{Nsum}) is a linear functional of ${\bf n}$,
the corresponding iso--$N$ surfaces $\Sigma_N$'s are just hyperplanes 
in $N^M$ (with $45^0$ inclination with respect to all $N$ coordinate axis).
Similarly, 
if the energy levels $\epsilon_c$'s in (\ref{Esum}) are 
independent on ${\bf n}$,
the $(M-1)$-dimensional iso--$E$ surfaces 
and the $(M-2)$-dimensional iso--$N$--iso--$E$ surfaces 
are {\it flat hyperplanes} as well.
However, 
if the energy levels depend on one or more occupation numbers $n^c$'s,
the $\Sigma_E$'s and $\Sigma_C$'s will all be {\it curved surfaces}
(see the next Sect.\ref{MacroRiemCurv}).
Here, we talk of curvature in the most intuitive sense of the term, 
according to the distance defined by (\ref{dnn}).

Describing thermodynamical states in terms of 
$Q$--dimensional curved surfaces
embedded within an higher--dimensional space 
has been considered before by many
(\cite{Gilmore84};
\cite{Ruppeiner85},\cite{Horn85},\cite{Gilmore85};
\cite{SNB85},\cite{Levine86},\cite{MNSS90}).
There are differences with our approach; 
basically, 
(i) our states ``live'' in a much larger $M$--dimensional space
containing many more (also non-equilibrium!) states,
and containing as submanifolds
the $(Q+1)$--dimensional or $(2Q+1)$--dimensional
embedding spaces previously considered by such authors,
and/or
(ii) they metric of the embedding space 
is itself non--euclidean to start with.
We discuss this in Sect.(\ref{Discu}).

%%%%%%%%%%%
\subsection{Deforming surfaces: energy (pseudo-)levels \\
and induced (riemannian, curved) metrics}
\label{MacroRiemCurv}

In many cases
familiar from standard, extensive 
thermo--statistical--mechanics 
(the ideal gas, 
a system of spins interacting with an external magnetic field
but not with each other, etc.)
we may take the energy levels $\epsilon_c$'s
to be {\it constants}, known a priori
and independent on the occupation numbers $n^c$'s,
i.e. 
$\frac{\partial E}{\partial n^c}=\epsilon_c$ 
and
$\frac{\partial^2 E}{\partial n^b \partial n^c}=
 \frac{\partial \epsilon_a}{\partial n^b}=0$ 
for any $a,b$.
In this case
(\ref{Nsum}) and (\ref{Esum}) are linear functionals of ${\bf n}$,
so all the iso--$N$ surfaces, the iso--$E$ surfaces, and
their intersections $\Sigma_C$'s are just hyperplanes in $N^M$.
This is the case for the ideal gas, where 
$
\label{levelkin}
\epsilon^{v}_c=\frac{1}{2} m {\vec v}_c
$
are purely kinetical energy levels.
Similarly,
for a system of particles interacting with an {\it external} potential,
we have 
will be the ``external potential energy levels''
$
\label{levelext}
\epsilon_c^{X}(X) = \phi^{(ext)}_c(X)  
= \phi^{X}({\vec r}_c,{\vec v}_c;X)
$,
depending on some external parameter(s) $X$
-- say, the component(s) of an external magnetic field,
if the system is made of electrically charged particles.
In both cases,
we are dealing with single--particle energies,
like those felt by a test particle 
interacting only with a fixed potential.
In fact, regarding velocity as a coordinate in $\mu$--space,
kinetic energy per unit mass may also be regarded as a fixed, 
$\mu$--space ``kinetic potential''.

However, in general the energy levels {\it themselves} 
might depend on the mesostate ${\bf n}$ of the system.
Physically, this is very relevant:
we would like to consider interactions {\it among} particles,
at least a pairwise interaction $\psi_{ij}$. 
But -- anytime the interaction are long(enough)--range,
i.e. when the range is larger than the side of the cell --
this translates into a nonlinear interaction energy 
$n^c \psi_{ab} n^b$ among particles in the $c$-th and $h$-th cells.
We may rephrase such a situation by introducing 
the ``interaction energy levels'' 
$
\label{levelint}
\epsilon^{\psi}_c({\bf n})=\frac{1}{2}\sum_b \psi_{bc} n^b
$
associated with the $c$-th cell only, but depending on all $n^b$'s;
the factor $1/2$ compensates for counting twice the same pair,
once as $(b,c)$ and once as $(c,b)$.
In general in (\ref{Esum}) we must insert the total energy levels: 
\begin{equation}
\label{leveltot}
\epsilon_c({\bf n};X) 
= \epsilon^{v}_c + \epsilon^{X}_c(X) + \epsilon^{\psi}_c({\bf n}) 
\end{equation}
Once more, such physical statements as
readily translate into a coincise geometrical language.
The local ${\bf n}$ dependence of $\epsilon_c({\bf n};X)$ means 
the iso--$E$ surfaces are not just flat hyperplanes,  
but rather $(M-1)$--dimensional {\it curved} hypersurfaces;
the overall dependence on $X$ means there exist an external
``handle'' with which we could globally ``tune'' the value of curvature.
The local, variable ``slopes'' of $\Sigma_E$ at ${\bf n}$
are given by the local ``pseudolevels'':
\begin{equation}
\label{pseudoleveltotsum}
\varepsilon_c^{tot}({\bf n};X) 
	:= \frac{\partial E}{\partial n^c}({\bf n};X) 
	= 
{\left[ 
\epsilon_c+ \sum_{b=1}^M n^b \frac{\partial \epsilon_b}{\partial n^c}
\right]}({\bf n};X) 
\end{equation}
or, in a more compact geometric notation, by:
\begin{equation}
\label{pseudoleveltotscalar}
{\underline \varepsilon_c^{tot} ({\bf n};X) }
=
{\left[
{\underline \epsilon_c}+\left( {\bf n} \cdot {\underline \nabla} \right)
{\underline \epsilon_c}
\right]}({\bf n};X) 
\end{equation}
where ${\underline \nabla}$ is the covariant gradient operator
w.r.t. the euclidean coordinates $n^1,..n^M$.

Now, 
since any $\Sigma_E$ is anyway a metric submanifold embedded within 
the euclidean manifold $(N^M,{\underline {\underline \delta}})$,
the {\it metric} (co)tensor ${\underline {\underline g}}^{(E)}$ induced by 
the metric $2$--form ${\underline {\underline \delta}}$ onto $\Sigma_E$
is also positive--definite (=all eigenvalues of $g_{ab}^{(E)}$ are positive):
no matter where they are,
two different mesostates have always non--zero, positive distance.
However, the components $g_{ab}^{(E)}$'s of the induced metric tensor 
are different from the components $\delta_{ab}$'s of the 
ambient metric tensor; in particular, they are 
``position''--dependent functions $g_{ab}^{(E)}({\bf n})$.
Indeed,
if 
we define an $(M-1)$--dimensional surface $\Sigma_A \in N^M$ 
by setting $A({\bf n})=constant$, 
and if 
along the $c$--th direction we have 
${\bf e}_c \cdot {\underline \nabla A} \not = 0$,
then 
the $(M-1) \times (M-1)$ components $g_{ab}^{(A)}$
(with $a,b=1,...M$ but $a,b\not = c$)
of the metric ${\underline {\underline g}}^{(A)}$
induced upon $\Sigma_A$
will read:
\begin{equation}
\label{gabA}
g_{ab}^{(A)}({\bf n})
 = \delta_{ab} +
{\left[
\frac{
\frac{\partial A}{\partial n^a}
\frac{\partial A}{\partial n^b}
}{
\left( 
\frac{\partial A}{\partial n^c}
\right)^2
}
\right]}({\bf n})
\end{equation}
which is manifestly ${\bf n}$--dependent.

Does this position--dependence of the metric coefficients
mean the iso--$A$ surfaces are curved?
No, if we were allowing for more general coordinates;
a plane is still flat even when described in polar coordinates,
whose metric coefficient depend on position
through the Jacobian of the cartesian--polar transformation
of coordinates.
Indeed yes, as we are here using cartesian and orthogonal coordinates.
In other words, though the embedding cell space is flat and euclidean,
the curvature of the $\Sigma_A$'s still depends on their shape.
This can only be checked for by computing the 
{\it curvature} tensors (scalar, Ricci, Riemann or Weyl, etc.)
%-- see Appendix B)
%\ref{AppGeom}),
which depend on the metric ${\underline {\underline g}}$
{\it and its $1$--st and $2$--nd order derivatives} w.r.t. the $n^c$'s
(position--dependence in principle only requires 
$\frac{\partial g_{ab}}{\partial n^c} \not =0$ 
for at least one choice of $a,b,c$, a much weaker condition than curvature).
Such tensors could be negative-- as well as positive--defined or undefined
-- again, something that can be checked explicitly and straightforwardly,
once given ${\underline {\underline g}}({\bf n})$.

Note how curvature arise due to particle--particle interactions.
Such a connection 
has been first pointed out by 
Ruppeiner (\cite{Ruppeiner79},\cite{RuppeinerRev95}).
However, 
(i) his curvature is computed starting from
a different (entropy--based) definition of distance,
and
(ii) he adopts an {\it intrinsic} point of view,
where one needs in principle no embedding space
to define a metric onto $\Sigma_A$,
while we use an {\it extrinsic} point of view,
where one needs an embedding metric space ``all around'' $\Sigma_A$, 
to induce a metric onto it.
We discuss this in Sect.(\ref{Discu}).

%%%%%%%%%%%
\subsection{External/kinetic potentials vs pairwise interactions:
linear vs non-linear transformations?}
\label{MacroExt/Kin/Int}

The effect of the modifying the external physical parameter $X$
can be geometrically understood as follows.
The external field $\phi^{ext}_c(X)$
depends on the cell label $c$ through $({\vec x}_c;{\vec v}_c)$,
but it does not depend on the occupation number $n^c$.
Every $c$ corresponds to a single direction in cell space.
Changing $X$ will modify 
the external potential energy levels $\epsilon_c^{X}(X)$,
in a fashion depending on the ``direction'' $c$, 
but {\it not} on the ``position'' $n^c$.
In geometrical terms, 
this operation is not isotropic, but still homogeneous.
However, since $E_{\phi}$ is a linear functional of the $n^c$'s
it makes sense to relate the old external energy levels to the new ones by
some suitable coefficient $c_c(X \to X')$, independent on ${\bf n}$.
Correspondingly, the total energy levels will change to
\begin{equation}
\label{levelexttransf}
\epsilon_c({\bf n}; X') = 
\epsilon_c({\bf n}; X)+[c_c(X \to X') - 1] \epsilon_c^{X}(X)
\end{equation}
Geometrically, 
this effect can be viewn as a {\it linear transformation}
of the set of iso--$E$ hyperplanes in cell space,
induced by $X \to X'$.
This transformation is homogeneous, but not isotropic.
Furthermore, 
since different directions do not get ``mixed up'' with each other,
the linear transformation (\ref{levelexttransf}) is simply 
a composition of a differential dilatation/contraction with a translation,
but without any rotation involved.

Note how the total kinetic energy
$E_{k}({\bf n})=\sum_c n^c \epsilon_c^{v}$
and the total external potential energy 
$E_{\phi}({\bf n})=\sum_c n^c \epsilon_c^{X}(X)$
are both {\it linear} functional of ${\bf n}$,
while the total 2--particle interaction energy 
$E_{\psi}({\bf n})=\sum_c n^c \epsilon_c^{\psi}({\bf n})$
is a {\it nonlinear} functional (of order 2 in the $n^c$'s) of ${\bf n}$.
Now, 
the transformation (\ref{levelexttransf}) involving $E_{\phi}$
turns out to be linear.
This suggests that there might be some other linear transformations
associated with $E_{k}$, 
and some {\it non linear transformation} (of order 2?)
associated with $E_{\psi}$, 
in general inhomogeneous and/or anisotropic
(e.g., as an hyperbolic paraboloid).
 We might pursue this suggestions one step further.
 Transformations in general
 are associated with invariance and symmetry properties,
 in turn associated with physical conservation laws (Noether's theorem).
 We have just speculated that there might be a linear transformation
 associated with kinetic energy levels; it would be natural
 to expect this postulated transformation be associated
 with the conservation of kinetic energy upon 
 linear transformation of the (physical) coordinate system.
 We might then speculate that the postulated nonlinear transformation
 associated with the interaction energy
 could be associated to some nonlinear transformation of our
  position/velocity coordinate system.
%In the case of a power--law potential $\psi(r) \propto r^\alpha$,
%it would be natural to expect such a transformation
%to be invariant upon a power--law rescaling of positions and velocities.
%Physically, 
%that would mean that it might be possible to map
%different thermodynamic systems or mesostastes
%(this is unclear as yet) onto each other,
%and regard them as in some (yet unclear) sense as 
%``thermodynamically equivalent''.

%%%%%%%%%%%%%%%%%%%%%%%%%%%%%%%%%%%%%%%%%%%%%%%%%%%%%%%%%%%%%%%%%%%%

\section{Equilibrium and entropy}
\label{Entro}

%%%%%%%%%%%
\subsection{Sequences of thermodynamic equilibria:\\ 
(curved) trajectories in cell space}
\label{EntroEquiTraj}

A paradigmatic situation encountered in 
thermo--statistical--mechanics is the following.
First, 
we consider a thermodynamical system
whose macrostates can be {\it completely} described by
a few macroscopic parameters -- say, $N$ and $E$ --
variable at will by the experimenter.
Second,
we are given the corresponding mesoscopic functionals
$N({\bf n})$ and $E({\bf n})$.
Third,
we are told the numerical values $N^*$ and $E^*$ 
assumed by $N$ and $E$, respectively.
Finally, 
we are instructed that, among all possible mesostates
${\bf n^*}', {\bf n^*}'',{\bf n^*}''',...$
compatible with the imposed $N^*$ and $E^*$,
the system will ``tend to select'' (somehow guided by the physics)
{\it one particular mesostate} $\hat {\bf n}^*=\hat {\bf n}(N^*,E^*)$,
depending on the imposed constraints 
$N(\hat {\bf n}^*)=N^*$ and $E(\hat {\bf n}^*)=E^*$.
Varying $N$ and $E$ will make $\hat {\bf n}(N,E)$
``wander around'' in cell space, 
along a 2--dimensional ``equilibrium trajectory'', 
or equilibrium surface, $\hat \Sigma(N,E)$.

The shape of the equilibrium trajectories depends on the
``convolution'' between the chosen coarse--graining 
and the underlying physics of the system.
For instance, it might happen that varying only $N$ but not $E$,
the equilibrium occupation numbers $\hat n^c$'s change
but the equilibrium probabilities $\hat p^c$'s do not.
This is the well--known case of the ideal gas,
where $\hat n^c \propto N \exp{(-\beta \epsilon_c^{v})}$
and $\frac{\partial \epsilon_c}{\partial N}=0$,
such that $\hat p^c \propto \exp{(-\beta \epsilon_c^{v})}$
does not depend on $N$. 
In other words, changing from $N$ to $N'$ 
the probability distribution remains the same 
${\bf p}={\bf n}/N={\bf n}'/N'$.
Geometrically, this is like multiplying ${\bf n}$ by $x=N'/N$,
i.e. ${\bf n}(N \to N', E)$ will ``move'' 
along a {\it straight line} of direction ${\bf u}_n=N {\bf p}/{\bf |n|}$
passing through the origin (the completely empty mesostate) of $N^M$.
Once again, 
for a non--ideal gas where interactions are not neglectable, 
the number of particles will be important, and
equilibrium states of different $N$
will lay along {\it curved trajectories} in $N^M$ space.

%%%%%%%%%%%
\subsection{States of thermodynamic equilibrium:\\ 
constrained extrema of the entropy functional}
\label{EntroExtr}

The way according to which the thermodynamical systems
``chooses'' $\hat {\bf n}^*$ among all candidate mesostates
${\bf n}^* \in \Sigma_C^*$ will be dictated by its physics.
However, as often happens,
it might be possible to mathematically rephrase this choice
in terms of a variational principle,
the {\it Maximum Entropy} principle (\cite{Jaynes57}).

First of all, one is provided with another,
given functional of ${\bf n}$, the entropy functional $S(...)$.
The physically outstanding example of such an entropy functional
is the usual $1$--particle Boltzmann--Shannon entropy $S_B({\bf n})$
of standard statistical thermodynamics and information theory
(\cite{Jaynes57}):
\begin{equation}
\label{SBsum}
S_B({\bf n}) 
	= N  \sum_{c=1}^M p^c \log (1/p^c)
	=    \sum_{c=1}^M n^c \log \left (\frac{N({\bf n})}{n^c} \right)
\end{equation}
This is appropriate to describe a system of many, weakly interacting particles.
Other forms of entropy functionals, more suitable for other physical systems,
are currently receiving a lot of attention, see Sect.(\ref{Discu}).
Second, 
in order to identify the desired equilibrium state $\hat {\bf n}(N^*,E^*)$,
one is instructed to look for the (constrained) maximum of $S({\bf n})$
onto $\Sigma_C^*$. 
Right there, should lye the wanted $\hat {\bf n}^*$.

More insight is gained if the problem is formulated in geometrical terms.
We have already introduced a few different submanifolds of $N^M$: 
the level surfaces 	  $\Sigma_S(S)$, 
the constraint surfaces   $\Sigma_C(N,E)$,
and the equilibrium surface  $\hat\Sigma$.
They all have different dimensionality, 
and in general lye along completely different and variable directions in $N^M$.
Their intersections define further subsets of $N^M$.
The constraint surface $\Sigma_C^*$ cuts all through $N^M$, 
through the iso--entropy level surfaces $\Sigma_S$, 
and the extremum surface $\hat\Sigma$, as well.
The intersection
$\Sigma_{CS}^*(S)=\Sigma_C^* \cap \Sigma_S$ 
induces a new 1-parameter family of ``constrained entropy surfaces'' 
onto $\Sigma_C^*$.
Finally, 
the intersection
$\hat\Sigma_{N^*E^*} 
= \hat\Sigma \cap \Sigma_N^* \cap \Sigma_E^*
= \hat\Sigma \cap \Sigma_C^*
$ 
defines a point ${\bf n}^*=\hat {\bf n}(N^*,E^*)$
(``constrained extremum mesostate'')
{\it simultaneously} belonging to 
the extremum family {\it and} the constraint surface(s).

Clearly, 
once a partitioning of phase--space has been chosen,
the shape of the equilibrium surface depends on 
(i) the functional form of the entropy $S(...)$, and
(ii) the functional form of the constraints $N(...)$ and $E(...)$.
The precise location of the desired constrained equilibrium ${\bf n}^*$
will also depend on
(iii) the numerical values $(N^*,E^*)$.

%%%%%%%%%%%
\subsection{Mesoscopic vs macroscopic entropy}
\label{EntroMesoMicro}

Varying the numerical values of $N$ and $E$,
the set of all points $\hat {\bf n}$ 
spans a 2--dimensional submanifold of $N^M$,
the extremum surface $\hat \Sigma$. 
Correspondingly, 
we can define a function
$\hat S(N,E):= S(\hat {\bf n}(N,E))$,
i.e. the restriction of
$S({\bf n})$ to $\Sigma_C(N,E)$,
which we will call the {\it macroscopic entropy function},
as opposed to the {\it mesoscopic entropy functional} $S({\bf n})$ in $N^M$.

It is important to stress the conceptual difference among
$S({\bf n})$ and $\hat S(N,E)$. 
The former 
is a functional depending on the {\it mesoscopic} variables $n^1,...n^M$,
so it is always defined {\it everywhere} in $N^M$.
The latter, instead, 
has been introduced only after we focused attention on 
the particular family of mesostates $\hat{\bf n}(N,E)$ in $N^M$.
However, $\hat S(N,E)$ can now also be regarded 
as a function of the ${\it macroscopic}$ variables $(N,E)$ only.
In principle, $\hat S$ is then not related anymore
to any particular mesoscopic state of the system,
as long as ${\bf n}$ is compatible with the values of $N$ and $E$
as computed from (\ref{Nsum}) and (\ref{Esum}).
Of course, 
the two entropies coincide when they are evaluated 
{\it upon the extremum surface} $\hat\Sigma_{NE}$.

%%%%%%%%%%%
\subsection{Derivatives, metric(s), and curvature(s):
\\
a mini-review of previous work}
\label{EntroDerMetrCurv}

Standard thermo--statistical mechanics
is dominated by the notions of energy, entropy, and equilibrium,
which has had a big impact on its geometrical implementations,
as can be seen by the following review.
First, 
since Gibbs (\cite{Gibbs})
the geometrical meaning of {\it first} derivatives of $S$,
and the associated Legendre--transform structure
connecting intensive and extensive thermodynamic variables
has been fully appreciated.
Second, 
Weinhold (\cite{Weinhold}) 
used the matrix $D^2 E$ of second derivatives of $E$ 
w.r.t. the macroscopic variables $(N,S,...)$
to define a scalar product between {\it equilibrium} macrostates, used in turn
as a very practical tool for book--keeping and compact derivations
of standard themodynamical identities involving derivatives,
i.e. the Maxwell relations.
Conversely, Ruppeiner (\cite{Ruppeiner79})
used the matrix $D^2 S$ of {\it second} derivatives of $S$ 
w.r.t. the macroscopic variables $(N,E,...)$
to define a {\it distance}
(=metric tensor=first fundamental form of a surface)
between macrostates
-- in a manner totally analogous to (\ref{dS22}).
He interpreted his distance 
as a measure of (un-)likeliness of fluctuations around a
macroscopic equilibrium state.
Weinhold's and Ruppeiner's metric structures where later
recognized to be (conformally, and physically) equivalent
(\cite{SNI84}, \cite{Mrugala84}).
Many other authors since (see in \cite{RuppeinerRev95})
%\cite{SAGB80}, 
%\cite{SB83},\cite{CJ85},\cite{DL85},
%\cite{FAQS85},\cite{NSAQA85},\cite{Schlogl85},,\cite{DLR89},\cite{NS88})
were naturally 
led to adopt the Weinhold--Ruppeiner's metric to define distances, 
and to investigate into its physical significance.
Furthermore, Ruppeiner with his coworkers (see in \cite{RuppeinerRev95})
especially considered the physical meaning of the (Gaussian) curvature 
(roughly speaking, 
a kind of second derivative of the metric; App.B), 
%App.\ref{AppGeom}), 
pointed out its connection with interactions,
and interpreted its numerical value and physical units
as a measure of (interaction--induced) correlations among fluctuations.
He also introduced a hierarchical picture 
of fluctuations--of--fluctuations--of--fluctuations,
which naturally lends itself to be linked
to ideas from critical phenomena and (real--space) renormalization groups.
Third,
as a consequence of such an entropy--energy--oriented view,
they took for granted that $dS \geq 0$ should always hold
-- which is often not the case, 
as convincingly proven by Gross and coworkers
(\cite{GM97}, \cite{Gross98},\cite{GV99}, \cite{Gross00})
with theoretical as well as experimental evidence.
So the previous authors did not see part of the complications.
Fourth,
when considering curvature 
they ended up making (unproven, and probably unphysical) statements
concerning the {\it third} derivatives of $S$ (\cite{Gilmore84}).
To be more precise, one would expect the curvature to involve 
{\it fourth--order} derivatives of $S$, but
in fact if the metric itself is defined in terms of second derivatives of $S$,
then all terms containing second derivatives of the metric
cancel and disappear from the associated Riemannian curvature 
(\cite{Gilmore84},\cite{JM89}).
Gilmore 
(\cite{Gilmore84},\cite{Gilmore85}) fiercely contested such an attitude,
pointed out its drawbacks,  and 
supported (\ref{dS12}) as a definition of distance.
Self--consistently, he interpreted (\ref{dS22}) merely
as a definition of {\it curvature}
(=Riemann tensor=second fundamental form of a curved surface
=first extrinsic derivative of the metric tensor),
thus avoiding problems both with the sign of $d_{S2}^2$
and with higher--order derivatives of $S$.
%Gross and coworkers (????;????;?????), dealing spatially small systems,
%further supported Gilmore's view 
%and brought it within the short--range realm 
%(attractive/disruptive/any--range + a small system 
%= ``as if'' attractive/disruptive/long--range).

%--------------------------------------------------------------------------
%
\section{Discussion}
\label{Discu}

Here we contrast and connect our ideas with previous work.
Concerning geometry and statistical thermodynamics,
two basic issues must be addressed:
(1) Euclidean vs (pseudo-)Riemannian geometry, and
(2) macroscopic (``equilibrium'') thermodynamics
vs  mesoscopic (including ``non--equilibrium'') statistical mechanics.
From a geometrical point of view, 
issue (1) makes {\it a lot} of difference,
while (2) involves just a very natural generalization
(from $Q$=a few to $M$=many more dimensions and coordinates).
From a physical point of view, 
issue (1) needs a full discussion, 
but (2) is not too trivial as well.
A further issue (3) non--extensivity
explicitly involves the entropy,
and its non--extensive generalizations.

%%%%%%%%%%%
\subsection{Geometry and thermo--statistics 1:\\
coordinates, metric(s), curvature(s)}
\label{GeomTherm1}

Let us start with issue (1) - geometry.
As mentioned in Sect.\ref{Intro},
both Euclidean and Riemannian approaches 
have been put forward in the past, and came into conflict
in the late 1980's, in a dispute summarized ant to some extent solved
in (\cite{SNB85}), though not yet completely settled
(cfr. the comments in \cite{GV99}, \cite{Gross00}).
The key question is: 
what is the ``best'' definition of distance between thermodynamical states?
As usual with such ``best''--questions, the answer is:
it really depends on what one is interested in.
In fact, 
we all want two states ${\bf n'}$ and ${\bf n'}$
to be called ``distant'' if they are ``in some sense'' very different.
But precisely {\it in what sense} should they be different?
In particular, 
we might like to regard two states as very different (and hence very distant)
if (A) they ``look'' very different when we take a ``snapshot'' of them
(hence, a {\it static} concept of distance), 
or if (B) they ``keep'' very different when we try to ``transform''
one into the other
(hence, a {\it dynamic} concept of distance).

What is the ``best'' attitude?
It depends whether one is interested 
in describing the system as it is
(keywords: information, probability, occupation, etc.),
or 
in describing the system as it evolves
(keywords: transformation, stability, transition, etc.).
If one chooses (A), then a very natural distance is:
\begin{equation}
\label{dn2}
d_n^2 = \sum_{a=1}^M \sum_{b=1}^M dn^a dn^b
\delta_{ab}
\end{equation}
which coincides with our definition (\ref{dnn}).
If one chooses (B) istead, then a natural guess would be:
\begin{equation}
\label{dS12}
d_{S1}^2= \sum_{a=1}^M \sum_{b=1}^M dn^a dn^b
\frac{\partial S}{\partial n^a}({\bf n})
\frac{\partial S}{\partial n^b}({\bf n})
\end{equation}
which is strongly reminiscent of (\ref{gabA});
but almost as natural it would be:
\begin{equation}
\label{dS22}
d_{S2}^2= \sum_{a=1}^M \sum_{b=1}^M dn^a dn^b
\frac{\partial^2 S}{\partial n^a \partial n^b}({\bf n})
\end{equation}
%
%All such definitions share the complications of covariance
%(choice of coarse--graining partition).

%
A beautiful logical pattern starts to emerge.
It takes little to recognize that
$d_{S1}=d^1S=$1--st order variation of $S$,
and 
$d_{S2}^2=d^2S=$2--nd order variation of $S$.
Involving the {\it first} derivative of a state function, 
it is immediate to relate $d_{S1}$ 
to exact differentials, thermodynamic potential, and the like;
since the state function is the entropy,
$d_{S1}$ should have some connections 
to the $1$-st law of thermodynamics, and to extremum/equilibrium points.
Involving the {\it second} derivative of a state function, 
$d_{S2}$ should be related to 
convexity/concavity, minimum/saddle/maximum points -- 
and hence equilibrium, too --  but in particular to (in)-stability,
and therefore transitions;
since the state function is the entropy,
$d_{S2}$ should have some connections 
to the $2$-nd law of thermodynamics.

So (\ref{dn2}), (\ref{dS12}) , and (\ref{dS22}) look physically appealing.
But, are they geometrically meaningful distances? 
Definition (\ref{dn2}) 
yields a perfectly reasonable and intuitive notion of distance (obviously),
$d_{n}^2 \geq 0$ always, and $d_{n}^2 = 0$ only trivially.
Definition (\ref{dS12}) looks good, but it has some complications.
In fact,
two states can have very different occupation ${\bf n}' \not = {\bf n}''$,
and in most cases it will be {\it also} very difficult
to realize a transition ${\bf n}' \to {\bf n}''$, between them,
{\it yet} they can easily have the same entropy
$S({\bf n}')=S({\bf n}'')$, so in this case $d_{S1}^2=0$.
However, the associated metric is still positive (semi-)definite,
i.e. $d_{S2}^2 \geq 0$ in general.
So such a $1$--st order  entropy--distance $d_{S1}$
whose path integral gives the entropy variation $|\Delta S|$,
would be useful if one were interested in
{\it quasistatic, reversible} thermodynamical transformations 
between ${\bf n}'$ and ${\bf n}''$,
i.e. two states are regarded as faraway if 
(C) it takes a high entropic price (=heat, dissipated work, $1$--st law) 
to go from one to the other.
Furthermore, 
though entropy--based (\ref{dS12}) is in fact closely related 
to the entropy--independent (\ref{dn2})!
Indeed, let us consider a given iso--$S$ surface $\Sigma_S$.
Everywhere perpendicular to it,
there is the gradient (co)vector ${\underline \nabla} S$.  
Let us suppose that, at least at some special point ${\bf n}_\#$ 
$\Sigma_S$ is orthogonal the $c$-th of euclidan directions
${\bf e}_c$.
There, as a particular case of (\ref{gabA}),
we can compute 
the distance $(d_n^{(S)})^2$ induced by (\ref{dn2}) onto $\Sigma_S$,
and we get
Pythagora's theorem ``projected''
along ${\bf e}_c$ and upon $\Sigma_S$:
\begin{equation}
\label{dndS12}
\left( \frac{ d_{S1}^2 }{ ({\underline \nabla S})_c } \right)^2
+
\left( d_n^{(S)} \right)^2
=
d_n^2 
\end{equation}
Definition (\ref{dS22}) implies some further generalizations 
of the notion of distance, to that of ``interval'' or ``separation''
familiar from Einstein's Relativity Theory.
Indeed, 
by definition the occupation--distance will always be $d_n^2 \geq 0$,
while 
the $2$--nd order entropy--distance $d_{S2}$
can lead to 
either (a) $d_{S2}^2>0$, or (b) $d_{S2}^2=0$, or even (c) $d_{S2}^2<0$.
Strange as it may sound, the last case is not unphysical at all!
(Even in standard extensive thermodynamics, 
it is forbidden in a canonical or a grancanonical approach,
but it naturally arises in a microcanonical approach 
-- \cite{GM97}, \cite{Gross98},\cite{GV99}, \cite{Gross00})
In fact, physically such three cases may be linked to
stability, indifference, and instability w.r.t. perturbations 
${\bf n}' \to {\bf n}''={\bf n}'+\Delta {\bf n}$, respectively.
In other terms:
(a) the system ``does not want'' to go 
from ${\bf n}'$ to ${\bf n}''$, 
unless we ``pay a high price''(=a long and positive distance$^2$,
i.e. a ``space--like'' separation) to ``convince'' it;
(b) the system ``does not care'' whether it is 
in state ${\bf n}'$ or ${\bf n}''$, 
${\bf n}' \to {\bf n}''$ transitions are for free
(=a zero distance$^2$, i.e. a ``null--like'' or ``ligh--like'' separation), 
or 
(c) the system ``really wants'' to go 
from ${\bf n}'$ to ${\bf n}''$, 
and it is so ``convinced'' of this that it is even ``willing to pay us''
a high price (=a long but negative distance$^2$,
i.e. a ``time--like'' separation). 
The path integral of (\ref{dS22}) should correspond
to some measure of ``overall willingness'' of the system
to go through the specified 
(and in general {\it finite--time, non--reversible}) tranformation.
Physically, there might be a state which is 
{\it statistically} very attractive because of its entropy, 
but still the system might find it very hard to {\it dynamically} evolve 
to such a place, e.g. if the involved time--scales are exceedingly large.

In geometrical terms,
(\ref{dn2})  defines an Euclidean geometry,
(\ref{dS12}) defines a Riemannian geometry, and
(\ref{dS22}) defines a {\it pseudo-}Riemannian geometry.
In the first case, 
the metric tensor is the same everywhere,
distances are positive semi--definite,
total curvature is always zero,
intrinsic and extrinsic curvature
can take on any value as long as they add up to a zero total curvature.
In the second case,
the metric tensor changes from place to place,
distances$^2$ are still positive semi--defined,
but 
total, intrinsic, and extrinsic curvatures can all be positive/zero/negative
(as long as they satisfy the Gauss--Peterson--Codazzi 
consistency conditions w.r.t. the metric).
In the third case,
the metric tensor changes from place to place,
distances$^2$ are sign--indefined,
and 
total, intrinsic, and extrinsic curvatures are sign--indefined too.
The Euclidean/Riemannian controversy of the 1980's
contrasted the pro and con's of (\ref{dn2}) as opposed to (\ref{dS22}),
and it was 
concluded with ``it is even possible that....unified...common perspective''
(\cite{SNB85}).
In this paper, we hope we have finally clarified ``were the truth lies'':
as usual, a part on each side, 
and maybe even some on an additional side, e.g. (\ref{dS12})
-- but see also the end of Sect.(\ref{GeomThermNext}).

%
%All this involves rather basic notions of 
%differential geometry and surface theory,
%both traditional and relatively standard tools of mathematical physics.
%So why has it not been fully recognized long ago
%by ``geometro--thermo--statistical people''?
%In our opinion, the reasons are rather sociological in nature.
%
%Physically, instability and therefore $dS\leq 0$ 
%arises due to the .......... of the interactions.
%If they are short--range, however, ...........
%
%It has long been known by astrophysicists 
%(``attractive/disruptive/long--range people'')
%that real self--gravitating--systems can often be found
%in ``quasi--equilibrium'' (=very stable and long lasting) states
%similar to model--states whose Boltzmann--Shannon entropy is a {\it minimum}
%($d_{S2}^2 < 0$) in their $\mu$--space neighborhood,
%if one uses $S_B({\bf n})$ inside (\ref{dS22})
%(e.g.,\cite{?????}).
%
%This basic fact, however, 
%was not perceived by ``short--range people'' in the 1980's,
%and still continues to be a source of surprise and puzzlement for many
%(e.g., \cite{????}).

%%%%%%%%%%%
\subsection{Geometry and thermo--statistics 2:\\
probability/occupation, information, entropy}
%{vectors/covectors, macro/micro, equilibrium/non--equilibrium}
\label{GeomTherm2}

Let us now move on to issue (2) - thermo--statistics.
As we mentioned in Sect.(\ref{MicroMeso}) and Sect.(\ref{Macro})
the use of vectors and covectors to characterize
thermodynamical states and observables is not new
(\cite{SNB85}), (\cite{Levine86}), and (\cite{MNSS90}).
However, we must stress that 
our definitions/interpretations and those of such authors
differ under three respects:

(1)   we employ unnormalized occupation numbers $n^c$
      and actual, total values of the macroscopic observables,
      while they use normalized probabilities $p^c=n^c/N$
      and, correspondingly, expectation values per particle;

(2)   our sum is over all $M$ mesoscopic cells,
      while 
      theirs runs only over the $Q$ available macroscopic constraints
      (usually $Q<<M$);

(3)   more importantly -- and this is a key difference of our approach! --
      {\it we do not derive} our notion of distance from
      the $1$--particle Boltzmann--Shannon entropy functional,
      but {\it we introduce it directly} as a natural consequence 
      of our definition of mesostate as due to 
      a coarse--graining of $1$--particle phase--space.

Such remarks are closely connected with each other,
and this becomes particularly clear comparing our work 
to a remarkable paper by Levine 
(\cite{Levine86}; see also \cite{SNB85} and \cite{MNSS90}).
formally very close to ours.
The key point of divergence lays in the
{\it physical motivations} behind the choice of the metric.
Quoting Levine (his Sect.II.A; notation and stresses are ours), 
``in a statistical approach, the state of a system is given by 
a {\it normalized} probability distribution $p^c=p^1,...p^M$ 
over $M$ (mutually exclusive and collectively exhaustive) alternatives...in 
physics these $M$ alternatives are themselves {\it states of the system}''.
He introduces a metric in order to be able to define a 
scalar product (his Sect.II.B).
Later (his Sect.II.C) he argues that 
``the simplest motivation for the choice of the metric is that
any normalized probability distribution have a unit norm
$1=\sum_b p^b = \sum_a \sum_b g_{ab} p^a p^b$.
The choice $g_{ab}:=\delta_{ab} /p^a = \delta_{ab} /p^b$ satisfies''
the required defining conditions 
(single--valuedness, continuity, differentiability, symmetry;
positive--definiteness is not mentioned) for $g_{ab}$ to be a metric.
He comments (his note n.9) ``the metric {\it is a point function},
so that we have a Riemannian metric'' -- while in the Abstract,
he writes ``the metric does {\it not} depend on the state of the system''
and talks about ``Euclidean geometry'' and ``Euclidean space''.
In fact, 
computing the scalar product of two ``physical states''
(his Sect.II.F) he explicitly says that ``the space is Euclidean''.
Note that a ``physical state'' hereabove is {\it not}
identified by the previously considered probabilities $p^c$'s,
but rather by some re--normalized components
$\tilde p^c=\sqrt{g_{cc}}p^c=\sqrt{p^c}$,
i.e. ``probability amplitudes''.
Though this does not greatly affect the formalism,
it undoubtedly introduces a rather mysterious complication,
which is completely avoided in our approach.
Finally, 
Levine argues that ``the macroscopic constraint 
are all one really needs to know'', so in fact he is taking $M=Q$,
and merely extending an intrinsically macroscopic view 
(avocated especially by Ruppeiner) 
from $S(N,E)$ to a more general case such as $S=S(N,E,V,\vec J,...)$.
Now, all this 
($M$ ``elementary'' states, probabilities, amplitudes, normalization)
makes a lot of sense if one has three things in mind:
(i) quantum mechanics, (ii) extensivity, and (iii) information theory.
Justifying amplitudes becomes harder in classical mechanics;
and normalization is not very well--suited to non--extensive systems,
where by definition $N$ is expected to play a key r${\hat o}$le.
Indeed, Levine himself (his Sect.III.E) remarks: 
``the formal structure would appear, at least to this author
to be more satisfactory (certainly so in the quantal case). Indeed,
some of the simplicity of [his] Sect.II results from
{\it not imposing normalization at the outset}''.
In his Appendix he then considers non--normalized states,
and shows the connection with homogeneous functions of degree 1 
-- a typical signature of extensivity.
Finally, if $M>>Q$, maximizing $S$ as a function of the
$M$ parameters (most of which maybe unknown!) becomes impossible 
or computationally unfeasible, and the ``utilitarian'' approach 
of information theory looses much of its power.

We completely agree with most of Levine's remarks.
However, we are mostly interested in
(i) classical mechanics, (ii) non--extensivity, and
(iii) providing a conceptual tool to describe and gain insight into 
a physical situation, even those aspects of it
we might not directly have informations about.
So 
(1) we cannot rely upon $M$ intrinsically discrete quantum states,
but must implement them through the coarse--graining, and physically
justify the latter, 
(2) we want to keep track of $N$ as well as of ${\bf p}$
in our definition of ``physical state'',
and find it very natural and convenient to
combine them both into the occupation ${\bf n}=N{\bf p}$, and
(3) in this paper we are not so much concerned 
with the practical applications 
(where only $Q=$a few parameters are known) of our formalism,
but with matters of principle such as insight, logics, and foundations
(where we {\it might}, given enough money, technology,
computing and man--power, {\it but do not need} to really 
know all the $M=$many occupation numbers).

As an extra bonus of our attitude, 
we need not link our notion of distance to any entropy whatsoever,
and work with explicitly Euclidean structures from the very beginning.
In fact, Levine proceeds similarly, but at some points
it looks as if he is invoking a Riemannian point of view, instead.
In particular (his Sect.III), he shows explicitly
the connection between his Euclidean--looking geometrical construction 
and two Riemannian--looking derivations from either 
(i) the (micro/mesoscopic) 
statistical--mechanic and information--theoretic
Boltzmann--Gibbs--Shannon entropy $S({\bf p})$
or 
(ii) the (macroscopic) thermodynamical entropy $S(A_1,A_2,...A_Q)$.
On one hand, this sounds very comfortable
-- a deep connection between what's new and what was already known.
On the other hand, however, it should have sounded somewhat
surprising: how comes entropy is so {\it closely related} to a distance
which was defined {\it independently} of entropy?
We will provide an answer at the end of the next section.

%%%%%%%%%%%
\subsection{Geometry and thermo--statistics 3:\\
entropy, non--extensivity, metrics}
\label{GeomThermNext}

Within the context of Tsallis' non--extensive statistical thermodynamics,
one considers a generalization of the entropy.
But as we said, 
it is not clear fully to us {\it of which} entropy
is Tsallis' entropy functional the generalization,
especially whether of the $N$--particle Gibbs entropy $S_G$,
or of the $1$--particle Boltzmann entropy $S_B$,
{\it both} expressible as logarithmic functionals
of the relevant probability distribution.
However, 
as we do not deal neither with probabilities
nor with intrinsically $N$--body quantities,
we are lead to consider the following generalization of (\ref{SBsum}):
\begin{equation}
\label{STsum}
S_T({\bf n}) 
	= N \sum_{c=1}^M p^c \frac{(1/p^c)^\delta - 1}{\delta}
	=   \sum_{c=1}^M n^c \frac{1}{\delta}
\left[ \left( \frac{N({\bf n})}{n^c} \right) ^\delta - 1 \right]
\end{equation}
where we slightly modified the usual notation by defining $\delta:=(1-q)$.
For $\delta \to 0$, we get $S_T({\bf n}) \to S_B({\bf n})$.
As Tsallis' formalism is thought to be appropriate to describe 
systems characterized by long--range interactions/correlations,
we may anticipate the same for the $1$--cell entropy functional (\ref{STsum}).

Geometrically, 
the iso-entropy level surfaces of $S_B({\bf n})$ and $S_T({\bf n})$ 
are clearly different -- in absolute value, but also in shape. 
Even adopting the {\it same} constraints $N({\bf n})$ and $E({\bf n})$
will lead to different intersections between the 
$\Sigma_S$'s and the $\Sigma_C$'s, i.e. different extremum states.
Moreover, because of the {\it interactions},
the energy constraint appropriate such systems 
will {\it also} be different from a simple linear functional
of the $n^c$'s.
Indeed other forms of the constraints,
containing powers of the $n^c$'s, have been already proposed
in the literature on non--extensive statistical thermodynamics.
However, the interest has been mostly in the properties of
extrema under given constraints.
However, to such a purpose
{\it any} monotonically increasing transformation of $S_T({\bf n})$ 
will lead to an equivalently good entropy functional,
providing one is also willing to re--define the
mathematical parametrization (no physics involved!)
of the distribution function on 
$\frac{\partial S}{\partial N}$ and $\frac{\partial S}{\partial E}$.
In other words, 
the {\it value} of $S$ on the iso--$S$ surfaces will be modified,
but the {\it shape} of the $\Sigma_S$'s, 
and hence their intersections, location of maxima/minima/saddle--points,
direction of the gradients, and so on, will stay just the same.

Statistically,
$S_B({\bf n})$ is appropriate to describe systems composed of many, 
(almost) {\it independent} constituents.
We then expect that $S_B({\bf n})$ should be replaced by $S_T({\bf n})$
when {\it correlations} (maybe of some special type) become very important.
Physically,
correlations are induced by {\it interactions}.
And as we have seen, 
in these cases the mathematical form of the energy constraints also changes,
from linear to non--linear.
So we must {\it simultaneously} change
(i) the entropy functional $S(...)$, and
(ii) the energy functional $E(...)$.
No wonder then that the extremum mesostates are described 
by equilibrium distributions
different from the Maxwell--Boltzmann distribution.
More important, in our opinion at least, is the following question:
is there some kind of {\it self--consistency} condition
between $S(...)$ and $E(...)$?
This is a fundamental question. It seems to have not been yet addressed,
in the literature on non--extensive statistical thermodynamics.

Finally, the answer we promised 
-- leading to yet more questions in the next section.
The entropy functional used by 
(\cite{SNB85}), (\cite{Levine86}), and (\cite{MNSS90})
is formally exactly the same as $S_B({\bf n})$.
This bears two consequences.
First, only in such a case do we have a proove that
distances measured within $\hat \Sigma_{NE}$ 
according to (\ref{dS22})
or according to $D^2S(N,E)$ do actually coincide.
Second, 
the mathematical properties of the logarithm in $S_B({\bf n})$ 
are also at the root of Levine's proove that 
the curvature associated to $D^2S(N,E)$ is zero.
This depends on the euclidean embedding space,
but in equal part it depends on the (flat!) shape 
of the surfaces therein.
Indeed, {\it both} effects have the {\it same} origin,
and they are exactly what we should expect on physical grounds!
In fact, 
logarithmic Boltzmann entropy=ideal gas=no interactions/correlations=no curvature,
but also 
logarithmic Boltzmann entropy=factorized distributions=exact mean field=extensivity,
or still
logarithmic Boltzmann entropy=Boltzmann exponential=additivity 1--particle
energies, and additivity of the interaction energies 
at scales larger than the range of interactions.

%%%%%%%%%%%
\subsection{Geometry as a thermo--statistical tool:\\
some open questions, and temptative answers}
\label{GeomQuest}

We start from the question raised in the last section, namely:
is there (and what is it) a deep thermo--statistical
connection between non--extensivity on one hand, 
and  geometry on the other hand?
We strongly suspect so.

Then we recall the discussion on derivatives in 
Sect.(\ref{EntroDerMetrCurv}) and Sect.(\ref{GeomTherm1}),
and provide some temptative answers.
We have seen that 
(\ref{dS12}) is strictly related to the $1$-st law of thermodynamics
(``exchanged heat=temperature times variation of entropy''),
while
(\ref{dS22}) is strictly related to the $2$-nd law 
(``the entropy of the Universe tends to a maximum'', 
Clausius/Kelvin/Helmholtz?).
Is then (\ref{dn}) maybe somehow related 
to the $3$-rd law
(``there exist only one zero-temperature state'', Nernst; \cite{Reif}),
or to the $0$-th law
(``two systems in thermal contact eventually come into
thermal equilibrium''; \cite{Reif}),
or to the $4$-th law
(''in thermodynamics there are only intensive/extensive variables = 
0-th/1-st order homogeneous functions of $N$; \cite{Landsberg84})?
Probably not to the $3$-rd law:
(i) such a law is a statement about 
a single thermodynamical state ${\bf n}_0$
of one system, not something involving two states ${\bf n}'$ and ${\bf n}''$,
and
(ii) it can be formulated as $S({\bf n}_0)=0$, i.e a statement on the entropy,
which does not appear at all in (\ref{dn2}).
Probably  not to the $4$-rd law as well:
one one hand, we are talking about ${\bf n}$, and hence $N$;
on the other hand, we are talking {\it only} about ${\bf n}$,
and not of all other thermodynamic functions.
It is tempting to think about the $0$--th law,
but then we should somehow modify
our interpretation of ${\bf n}'$ and ${\bf n}''$
into ``state of system 1'' / ``state of system 2''.
(We are currently investigating such an issue.)

Finally, we note yet a few more interconnections.
(i) Ruppeiner has shown how using Riemannian geometry 
one can improve upon the classical theory of fluctuation,
on the $2$--nd order Gaussian approximation as well as 
on the full (but nevertheless exponential) theory.
He also succesfully used his approach to provide us
with a new theory of critical phenomena
(\cite{Ruppeiner83},\cite{Ruppeiner91},
\cite{RuppeinerRev95},\cite{Ruppeiner98};
\cite{DFLF84}).    
(ii) His theory contains derivatives of the entropy
up to $3$--rd order, as $4$--th order derivatives actually drop out;
(iii) $n$--th order derivatives of the entropy are usually
related to $n$--th order moments of the distribution function,
so as the Riemannian metric $D^2S(N,E)$ contains
informations about $\langle \Delta N^2 \rangle$,
$\langle \Delta N \Delta E \rangle$, etc.,
the Riemannian curvature from $D^2S(N,E)$ contains
informations about $\langle \Delta N^3 \rangle$
$\langle \Delta^N \Delta E^2 \rangle$, etc.;
(iv) distribution functions associated with Tsallis' entropy are symmetric, 
so they have vanishing skewness (=$3$--rd order moment),
yet they have non--gaussian kurtosis (=$4$--th order moment).
Now the question is -- what does this all mean?
In synthesis, we may rephrase our previous remarks as follows.
Standard thermo--statistical mechanics
only deals with $1$--st order (averages) 
and $2$--nd order (r.m.s. fluctuations);
standard information theory of un--correlated systems
elegantly and compactly captures this state of affairs in
its Legendre--transform and maximum--likelihood structures
(\cite{Jaynes57},\cite{Jaynes62}; see also \cite{PP97}).
Critical phenomena arise when the physically ``true''(=average) behaviour
and the most--likely(=mode) behaviour of a system radically depart
from each other -- and this is due to the distribution being 
asymmetric (=skewness) around the average.
Ruppeiner's theory elegantly and compactly captures this state of affairs in
his pseudo--Riemannian theory, which is actually $3$--rd order.
Our conjecture is then:
Tsallis theory represents the $4$--th order(=kurtosis) step 
in such a ladder of theoretical understanding.
Concerning its physical roots and meanings
in terms of well--defined statistical and mechanichal concepts,
we also have some temptative answers,
but they are still partially unclear to us as well,
so we leave them for further separated investigations.

%%%%%%%%%%%
%\subsection{Non--equilibrium and non--reversible \\
%statistical thermodynamics}
%\label{NeqNrev}
%\begin{itemize}
%\item non-equilibrium STD 
%(quiet approach to equilibrium $<->$ friction? $<->$ ``velocity'' flow field
%$<->$ vortical-rotational/potential-irrotational flow? in/compressible flow?;
%violent approach to equilibrium $<->$ inertia $<->$ turbulence? 
%$<->$ ``acceleration'' vector field 
%$<->$ gradient-irrotational-potential scalar field?)
%\item non--reversible STD 
%(non-integrable i.e. rotational flow field? dissipation i.e. friction? 
% memory/forgetness?)
%\item why should we care, after all? In other words,
%      what can we learn, or do, with all this stuff? (good question!)
%\end{itemize}
%--------------------------------------------------------------------------
%
\section{Conclusions}
\label{Conclu}

We presented, and physically justified, 
a new geometrical approach to thermo--statistical mechanics,
with completely new point of view and motivations compared to previous work.
We showed how in our perspective we can understand better
several issues raised in the past.
Moreover, 
we pointed out the many (and seemingly deep) connections
between two as--yet--unconnected areas of investigation
in thermo--statistical mechanics,
and came up with many new and interesting questions.
Geometrical insight points to a deep connection among such issues, 
and argues for even more -- and systematic! -- questions, 
e.g. 
what is the geometrical meaning of the $n$--th derivatives of $S$?
what is the its physical meaning?
is there a geometrical object providing an all--order interpretation?
This all loudly calls for $M$--dimensional visualization, 
thus leading to the use of many already available visualization techniques,
in turn leading to further insight, new ways of presenting results,
analyzing data, projecting experiments, writing codes and running simulations.
In brief, we presented a method.

We did not present concrete numbers or graphs, sharp--edged results
to be compared with those from other theories, simulations, or experiments.
That was not the purpose here.
The declared purpose was of methodological as well as pedagogical kind.
In such cases, {\it the method is the result}.
We dearly hope other investigators can profit of our work.

%--------------------------------------------------------------------------
%Acknowledgments

We would like to thank R.A.~Treumann, 
K.~Arzner, U.~Konopka,  R.~Sutterlin, and A.~Ivlev 
for a very nice time and many helpful discussions.
This work was supported by a post-doc ``Stipendium''
of the Max--Planck--Institut fuer extraterrestrische Physik, Garching,
where it was started. 
We also like to thank A.~Dominguez for many enligthening discussions
on coarse--graining and statistical mechanics,
and A.Wagner, T.Buchert, M.Kerscher, J.Schmalzing, and C.Beisbart
at Theoretische Physik, Ludwig--Maximillians--Universitaet, Muenchen.
%--------------------------------------------------------------------------

%--------------------------------------------------------------------------
%Bibliography
%

%--------------------------------------------------------------------------
%
\appendix{\bf Appendix: Definitions of entropy}
\label{AppEntro}

Here we give the definitions (with a few keywords)
of the many different function(al)s that in different contexts 
go under the same name of ``entropy''.

In a statistical--thermodynamical context, we are familiar with:
\par
\noindent
- the Clausius entropy 
(thermodynamic, macroscopic, Gibbs--Duhem):

$$
S_C(N,E,V)=\int (dQ +\mu dN - p dV)/T
$$ 
\par
\noindent
- the Boltzmann--Einstein entropy 
(probabilistic, macro-microscopic, fluctuation):
$$
S_\Omega=\ln \Omega(N,E,V)
$$
\par
\noindent
- the Boltzmann--Shannon entropy 
(kinetic, microscopic):
$$
S_B(N,{\bf p}_1)=-N\sum_c p_1^c \ln p_1^c \,\,\,\,\,\,\,(\mu-space)
$$ 
\par
\noindent
- the Gibbs--Shannon entropy 
(information theoretic, microscopic, Shannon--Wiener and Jaynes),
\par
\noindent
$$
S_G({\bf P}_N)=-\sum_c P_N^c \ln P_N^c   \,\,\,\,\,\,\,(\Gamma-space)
$$ 
Here
${\bf p}_1$ is the 1-body probability distribution function ($\mu$-space),
${\bf P}_N$ is the N-body probability distribution function ($\Gamma$-space),
$c$ is a label (not a power!) that identifies a suitably defined cell
in the relevant phase space,
$q$ and $\delta:=(1-q)$ instead are truly exponents,
$\Omega(N,E,V)$ is the $\Gamma$-space phase-volume compatible with 
total number of particles $N$, total energy $E$, total volume $V$,
and $(T,\mu,p)$ are the thermodynamic temperature, 
chemical potential, and thermodynamic pressure, respectively.

Apart from the factor $N$, 
$S_B$ and $S_G$ share the same logarithmic mathematical structure, 
namely that of the Shannon entropy:
$$
S_S({\bf p}) = -\sum_c p^c \ln p^c  
	     = \sum_c p ^c \ln \left(\frac{1}{p^c} \right)
$$ 
\par
\noindent
where now ${\bf p}$ is an unspecified probability distribution
in an abstract and general probability space.
However, in the context of information theory
several other forms have been proposed, among which:
\par
\noindent
- the ``multifractal'' Renyi entropy:
$$
S_R({\bf p}) = \frac{\ln \left( \sum_c (p^c)^q \right) }{1-q}
	= \ln \left( \sum_{c=1}^M p^c (p^c)^\delta \right)^\frac{1}{\delta}
$$
\par
\noindent
- the ``polynomial'' Behar  entropy:
$$
S_P({\bf p}) = \left( \sum_{c=1}^M p^c (p^c)^\delta \right)^\frac{1}{\delta}
$$ 
\par
\noindent
the ``nonextensive'' Havrda--Charvat--Daroczy--Tsallis entropy:
$$
S_{T}({\bf p}) 
 = \frac{\sum_c (p^c)^q -1}{1-q}
	= \sum_{c=1}^M p^c \frac{(1/p^c)^\delta - 1}{\delta}
$$
\par
\noindent
There are then two versions (Boltzmann--like and Gibbs--like)
for each of them.
%--------------------------------------------------------------------------
%Figures
%
%--------------------------------------------------------------------------
% The End
%
\end{multicols}
\end{document}